\documentclass[twocolumn]{aa}
\usepackage{graphicx}

\def\cs{{c_{\rm s}}}
\newcommand{\dfrac}[2]{\frac{\displaystyle{#1}}{\displaystyle{#2}}}

\begin{document}

\title{Molecular hydrogen emission from protoplanetary disks}

\author{H. Nomura\thanks{Now at Department of Earth \& Planetary
Sciences, Graduate School of Science and Technology,
Kobe University, Kobe 657-8501, JAPAN} \and T. J. Millar}
\offprints{H. Nomura, \\ \email{hnomura@kobe-u.ac.jp}}

\institute{School of Physics and Astronomy, The University of Manchester,
PO Box 88, Manchester M60 1QD, UK }

\date{Received 3 February 2005/ Accepted 21 April 2005}

\abstract{
We have modeled self-consistently the density and temperature profiles 
of gas and dust in protoplanetary disks, taking into account 
irradiation from a central star. Making use of this physical
structure, we have calculated the level populations of molecular
hydrogen and the line emission from the disks. 
As a result, we can reproduce the observed strong line spectra of 
molecular hydrogen from protoplanetary disks, both in the ultraviolet
(UV) and the near-infrared,  
but only if the central star has a strong UV excess radiation. 
\keywords{line: formation -- molecular processes -- radiative transfer
-- planetary systems: protoplanetary disks} 
}

\maketitle
%

\section{Introduction}

It has long been believed that infrared excesses over the stellar
photospheric emissions, often observed in the spectral energy
distributions of young stellar objects (YSOs), arise from dusty
circumstellar disks (Cohen 1974; Kenyon \& Hartmann 1987; Beckwith \&
Sargent 1993).
More direct evidence for  circumstellar disks has been found 
recently in the form of optically thick dust
lanes against the scattered light of the surrounding optically
thin nebulae at optical and near-infrared wavelengths (e.g., Burrows et
al. 1996; Stapelfeldt et al. 1998, 2003; Cotera et
al. 2001; Jayawardhana et al. 2002) and as a near-infrared image of
the scattered light of the disk itself (e.g., Fukagawa et al. 2004). 

Furthermore, thanks to recent high spectral resolution and high sensitivity
observations, it has become possible to detect various molecular line
emission from protoplanetary disks (e.g., Dutrey et al. 1997).
In particular, observation of molecular hydrogen line emission from the
disks is important because it directly traces gaseous masses of the
disks, which are connected with giant planet formation, without assuming
the dust-to-gas ratio or CO-to-H$_2$ ratio.
Pure rotational molecular hydrogen line emission was detected towards
YSOs and debris-disk objects with the {\it Infrared Space Observatory
(ISO)} (Thi et al. 1999, 2001a, 2001b), while no emission has yet been
observed 
with ground-based telescopes (Richter et al. 2002; Sheret et al. 2003;
Sako et al. 2005). Ro-vibrational molecular hydrogen line emission with
narrow line widths was detected towards some T Tauri stars (Weintraub et
al. 2000; Bary et al. 2002, 2003; Itoh et al. 2003). In addition,
fluorescent molecular hydrogen line emission in the ultraviolet (UV)
wavelength band was observed towards some 
classical T Tauri stars (Herczeg et al. 2002, 2004; Bergin et al. 2004).
However, at present there is no established model
for this emission which takes into account the
global physical structure of protoplanetary disks. 

Historically, molecular hydrogen emission has been observed towards
various kinds of 
astronomical objects, such as shock surfaces associated with star
forming regions, reflection nebulae illuminated by nearby massive stars,
planetary nebulae, supernova remnants, external galaxies, etc. (e.g.,
Beckwith et al. 1978; Brown et al. 1983; Hasegawa et al. 1987; Burton et
al. 1992), and studied theoretically, for example, under the conditions
of shock or photon-dominated regions (e.g., Black \& Dalgalno 1976;
Shull 1978; Hollenbach \& McKee 1979; Draine et
al. 1983; Pineau des Forets et al. 1986; Black \& van Dishoeck 1987;
Wagenblast \& Hartquist 1988; Sternberg 1988, 1989; Sternberg \&
Dalgalno 1989; Draine \& Bertoldi 1996). 
These studies have suggested the importance of physical condition of the
objects - density, temperature, and UV irradiation - for exciting
molecular hydrogen. 

Now, the physical structure of protoplanetary disks is thought to be
controlled by irradiation from the central star. Modelling the density
and temperature profiles of dust and gas in the disks has been
developed in more and more realistic ways as more detailed observational
data become available (e.g., Kusaka et al. 1970; Kenyon \& Hartmann 1987;
Chiang \& Goldreich 1997; D'Alessio et al. 1998; Nomura 2002; Dullemond
et al. 2002; Dullemond \& Dominik 2004; Kamp \& van Zaldelhoff 2001;
Gorti \& Hollenbach 2004; Glassgold et al. 2004; Kamp \& Dullemond 2004;
Jonkheid et al. 2004). 

In this paper, we have modeled the density and temperature profiles of
protoplanetary disks self-consistently, taking into account the
irradiation from the central T Tauri star. We have used these to investigate
the abundance and excitation of molecular hydrogen in the disk, and the
observational properties of the molecular hydrogen emission from the
disk. In the following section, we present the physical model of the
disk:- the density and temperature profiles of the gas and dust in the
disk on the assumptions of vertical hydrostatic equilibrium and
local thermal and radiative equilibrium. In Sect. 3, we calculate
the abundance and level populations of molecular hydrogen, assuming 
statistical equilibrium among the levels. Making use of these physical
and chemical
profiles, we compute molecular hydrogen emission from the disk at 
infrared and ultraviolet wavelengths and compare with the
observations in Sect. 4. 
Finally, the results are summarized in Sect. 5.

\section{Physical model}

We consider an axisymmetric disk surrounding a central star with the
physical parameters of typical T Tauri stars; a mass of
$M_*=0.5M_{\odot}$, a radius of $R_*=2R_{\odot}$, and a temperature
of $T_*=4000$K (e.g., Kenyon \& Hartmann 1995).

\subsection{Basic equations for the disk structure}

The gas temperature and density distributions of the disk are obtained
self-consistently by iteratively solving the equations for hydrostatic
equilibrium in the vertical direction and local thermal balance between
heating and cooling of gas. The equation for vertical hydrostatic
equilibrium is given in cylindrical coordinates $(x,z)$ by
\begin{equation}
\dfrac{dP}{dz}=\cs^2\dfrac{d\rho}{dz}=-\rho g_z, \label{eq.2-1}
\end{equation}
where $P$ and $\rho$ represent the pressure and density, respectively.
The sound speed $\cs$ is defined as $\cs^2\equiv dP/d\rho=kT/m_{\mu}$,
where $k$ and $T$ represent Boltzmann's constant and the
gas temperature, and the mean molecular mass $m_{\mu}$ is set to be
$m_{\mu}=2.3m_{\rm H}$ ($m_{\rm H}$ is the hydrogen mass). 
The vertical gravitational force is set
$g_z=GM_*z/(x^2+z^2)^{3/2}$, where $G$ is the gravitational constant.
The condition, $\int_{z_{-\infty}}^{z_{\infty}}\rho(x,z)dz=\Sigma(x)$,
is set to be satisfied, where $\Sigma(x)$ is the surface density
at a radius $x$ defined below. We put $\rho(x,z_{\infty})=1.67\times
10^{-20}$  g cm$^{-3}$ for the boundary condition. 
The equation for the detailed energy balance at each point in the
disk is given by
\begin{equation}
\Gamma=\Lambda, \label{eq.2-2}
\end{equation}
where $\Gamma$ and $\Lambda$ are the sum of the relevant gas heating and
cooling rates, respectively (see Appendix A for details of the
heating and cooling processes).

The dust temperature profile plays an important role in determining
the disk structure because the gas temperature is well coupled with the
dust temperature near the midplane of the disks where the density is
high enough for efficient collisions between gas and dust particles (see
Sect. 2.4). 
In this paper we obtain the dust temperature profile by assuming
local radiative equilibrium between absorption and reemission of
radiation by dust grains at each point in the disk in spherical
coordinates $(R,\Theta)$,
\begin{displaymath}
4\pi\int_0^{\infty} d\nu \kappa_{\nu}(R,\Theta)B_{\nu}[T_d(R,\Theta)]
\end{displaymath}
\begin{equation}
\hspace*{1cm}=\int_0^{\infty} d\nu \kappa_{\nu}(R,\Theta)\oint d\mu d\phi I_{\nu}(R,\Theta;\mu,\phi), 
\label{eq.2-3}
\end{equation}
where $\kappa_{\nu}$, $I_{\nu}$, $B_{\nu}(T_d)$, and $T_d$ represent the
monochromatic absorption coefficient, the specific intensity, the Planck
function for blackbody radiation at a frequency $\nu$, and the dust
temperature, respectively. 
Local thermodynamic equilibrium, $\eta_{\nu}=\kappa_{\nu} B_{\nu}(T_d)$,
is assumed, where $\eta_{\nu}$ is the monochromatic emissivity.
The specific intensity is calculated by solving the axisymmetric
two-dimensional radiative transfer equation, 
\begin{displaymath}
I_{\nu}(R,\Theta;\mu,\phi)=\int_0^s \kappa_{\nu}(R',\Theta')\rho(R',\Theta')
\end{displaymath}
\begin{equation}
\hspace*{3cm}\times B_{\nu}[T_d(R',\Theta')]e^{-\tau_{\nu}(R',\Theta')}ds', 
\label{eq.2-4}
\end{equation}
where $\tau_{\nu}(R',\Theta')$ is the specific optical depth from
a point $(R',\Theta')$ to $(R,\Theta)$, by means of the short 
characteristic method in spherical coordinates (Dullemond \& Turolla
2000). Here we neglect the effect of scattered light (cf. Dullemond \&
Natta 2003).
As heating sources we consider the radiative flux produced by 
the viscous dissipation ($\alpha$-model) at the midplane of the disk,
$F_{\rm vis}=(9/4)\Sigma\alpha\cs_0^2\Omega$, and the irradiation from
the central star $F_{\nu, {\rm star}}$ (Appendix C; see also Nomura 2002
for details of the model). 

The surface density distribution of the disks, $\Sigma(x)$, is determined
based on the standard accretion disk model (e.g., Lynden-Bell \& Pringle
1974; Pringle 1981), by equating the gravitational energy release
of accreting mass to thermal heating via viscous dissipation at the disk
midplane at each radius, $x$, 
\begin{equation}
\dfrac{9}{4}\Sigma\alpha\cs_0^2\Omega=\dfrac{3GM_*\dot{M}}{8\pi x^3}
\biggl[1-\biggl(\dfrac{R_*}{x}\biggr)^{1/2}\biggr],
\label{eq.2-5}
\end{equation}
where $\cs_0$ and $\Omega=(GM_*/x^3)^{1/2}$ represent the sound speed at
the midplane and the Keplarian frequency, respectively. Here, we adopt
$\alpha=0.01$ for a viscous parameter, and assume that the disk has a
constant mass accretion rate of $\dot{M}=10^{-8}$ M$_{\odot}$ yr$^{-1}$.
The disk mass between the inner radius $r_{\rm in}=R_*=2R_{\odot}$ and the
outer radius $r_{\rm out}=100$AU is then $6\times 10^{-3} M_{\odot}$.

\subsection{Ultraviolet radiation fields}
Ultraviolet (UV) radiation fields affect the gas temperature of the
protoplanetary disks through the grain photoelectric heating (Appendix
A.1) as well as the photodissociation and photoionization processes
(Appendix B and Sect. 3). 

\begin{figure*}
\centering
\includegraphics[width=8cm]{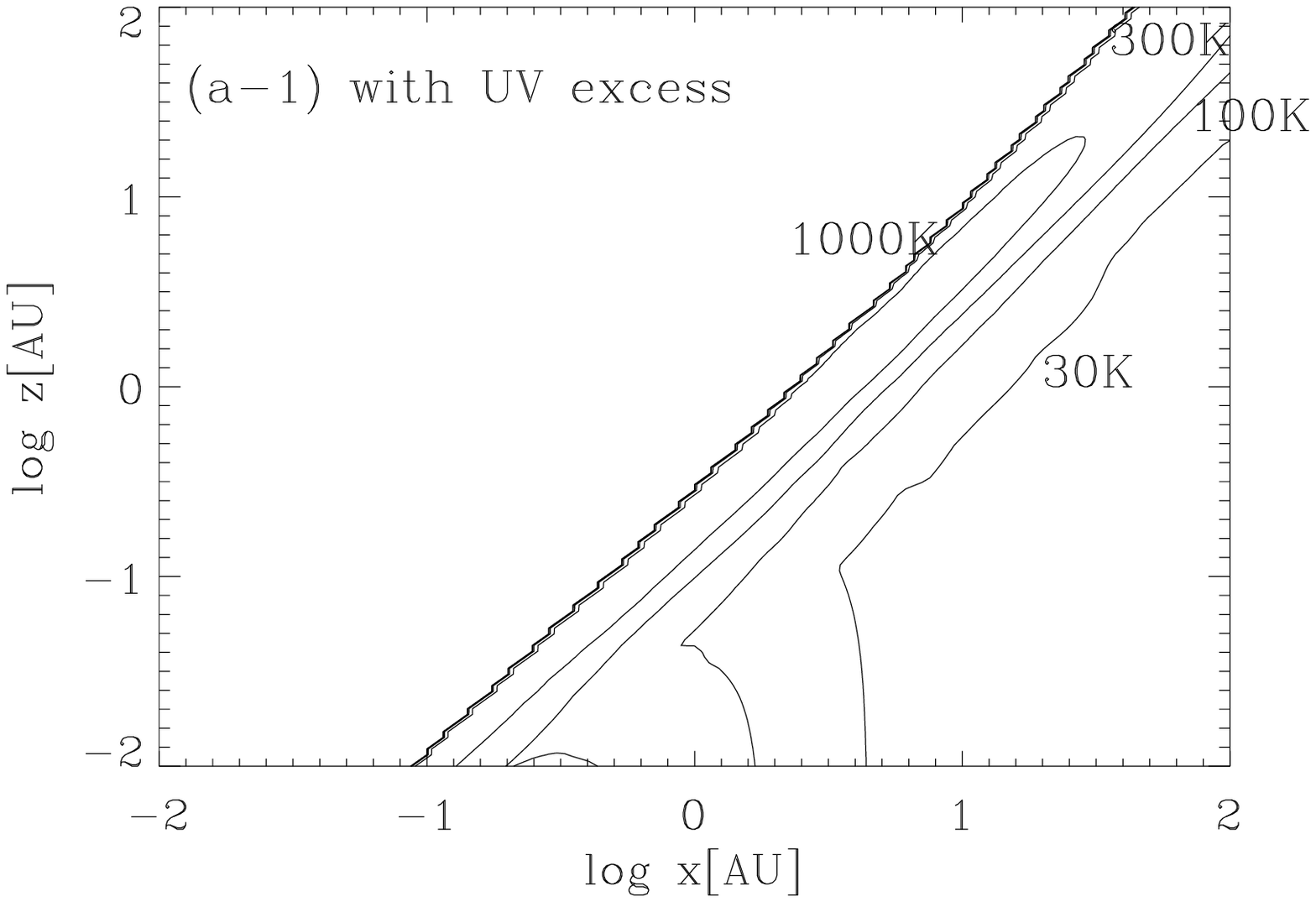}
\includegraphics[width=8cm]{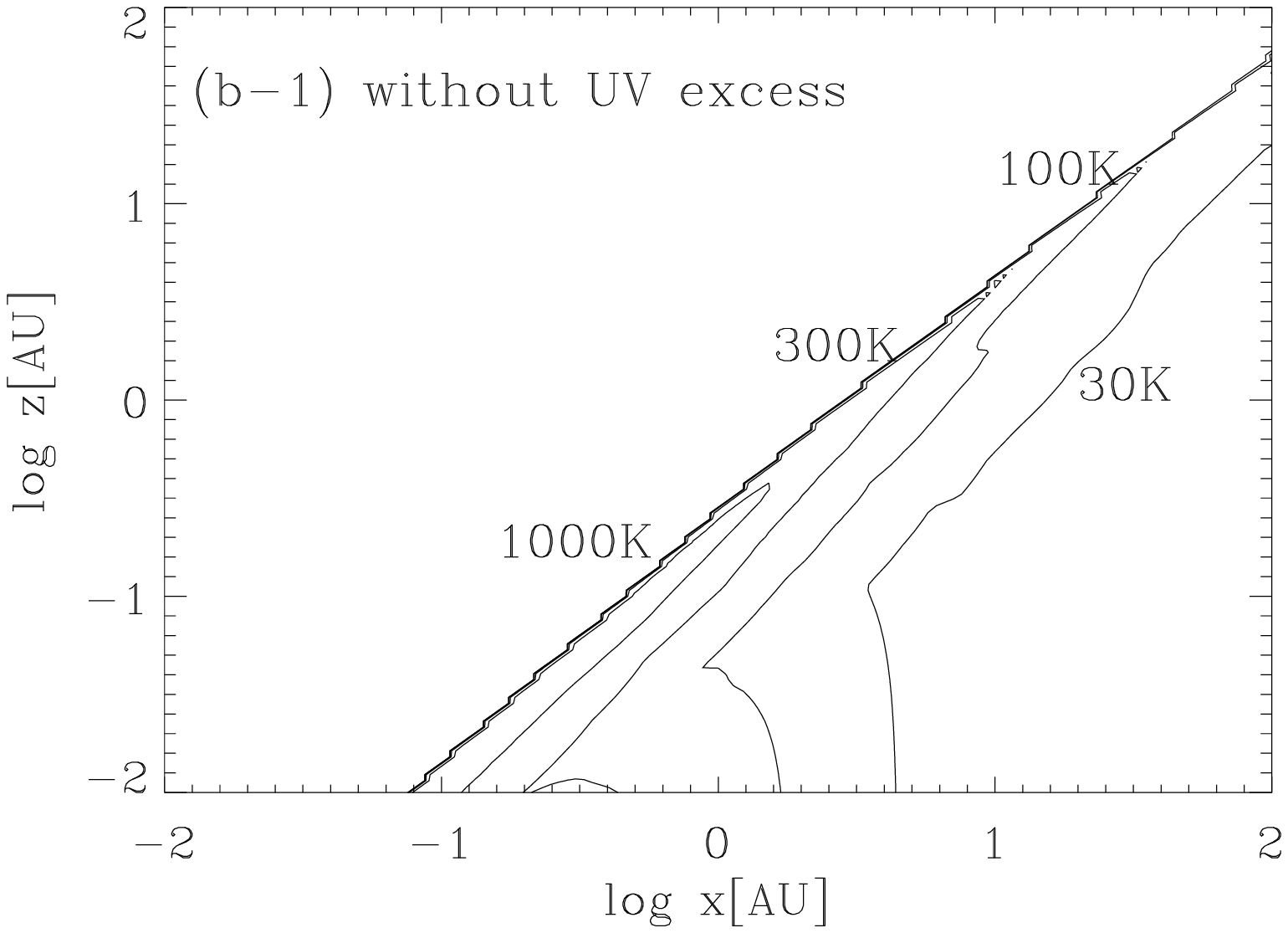}
\includegraphics[width=8cm]{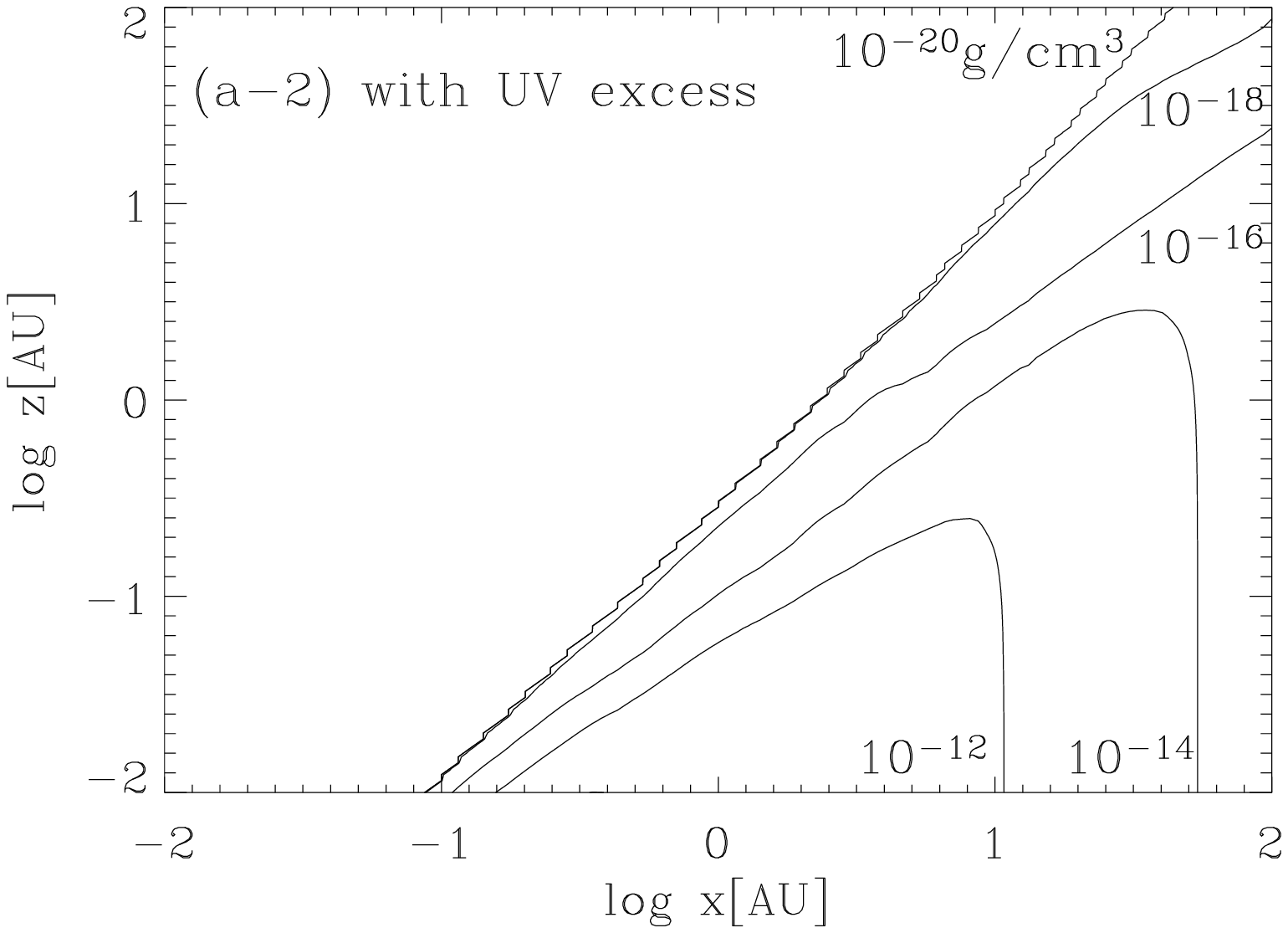}
\includegraphics[width=8cm]{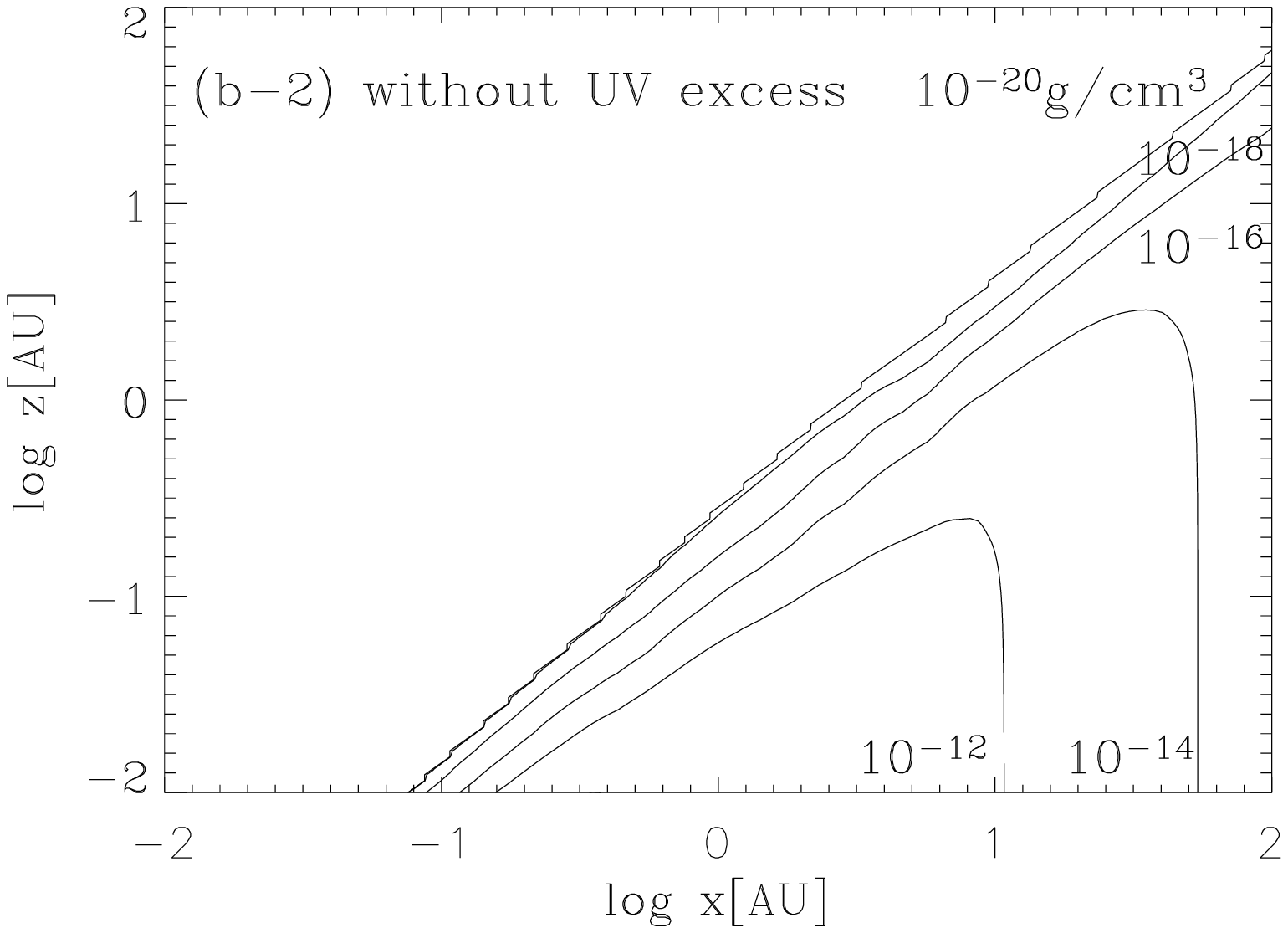}
\caption{{The gas temperature (a-1) (b-1) and density (a-2) (b-2)}
 distributions in the $x$-$z$ plane for the models (a) with and (b)
 without UV excess radiation of the central star. }
\label{f2.1}
\end{figure*}

The main UV radiation sources are generally thought to be due to
the central star and the interstellar radiation field. In this
paper we simply adopt a 1+1 dimensional UV radiation field.
In the radial direction, we calculate the radiation field which
directly comes from the central star as 
\begin{equation}
F_{\nu, R}(R,\Theta)=fF_{\nu, {\rm star}}\exp(-\tau_{\nu, R}),\ \ \ \tau_{\nu, R}=\int_{R_*}^R \chi_{\nu}\rho dR, \label{eq.2-6}
\end{equation}
where $F_{\nu, {\rm star}}$ is the specific radiation field at the
stellar surface, $f=(\pi/4)(R_*/R)^2$ the reduction of
radiation by a geometrical effect. $\tau_{\nu, R}$ is the specific
optical depth from the stellar surface $(R_*,\Theta)$ to a point
$(R,\Theta)$, and $\chi_{\nu}$ is the monochromatic extinction
coefficient defined by the absorption ($\kappa_{\nu}$) and scattering
($\sigma_{\nu}$) coefficients as
$\chi_{\nu}\equiv\kappa_{\nu}+\sigma_{\nu}$. In the vertical direction,
we consider the radiation source of the interstellar radiation field,
$F_{\nu, {\rm ISRF}}$, plus the contribution of the scattering of
radiation from the central star, 
\begin{displaymath}
F_{\nu, z}(x,z)=F_{\nu, {\rm ISRF}}\exp[-\tau_{\nu, z}(z_{\infty})]
\end{displaymath}
\begin{displaymath}
\hspace*{0cm}+2\pi\omega_{\nu}\int_z^{z_{\infty}} \kappa_{\nu}(x,z')\rho(x,z')F_{\nu, R}(x,z')e^{-\tau_{\nu, z}(z')}dz',
\end{displaymath}
\begin{equation}
\tau_{\nu, z}(z')=\int_z^{z'} \chi_{\nu}\rho dz'', \label{eq.2-7}
\end{equation}
where $\omega_{\nu}$ is the monochromatic albedo defined by the
absorption and extinction coefficients as
$\omega_{\nu}\equiv\sigma_{\nu}/\chi_{\nu}$, and $\tau_{\nu, z}(z')$ is
the specific optical depth from a point $(x,z)$ to $(x,z')$.
The approximate treatment of the scattered light in this model could
overestimate the UV radiation fields in the disks. Fully
two-dimensional radiative transfer calculation including the scattering
process should be done in future (e.g., van Zadelhoff et al. 2003). 

\begin{figure*}
\centering
\includegraphics[width=8cm]{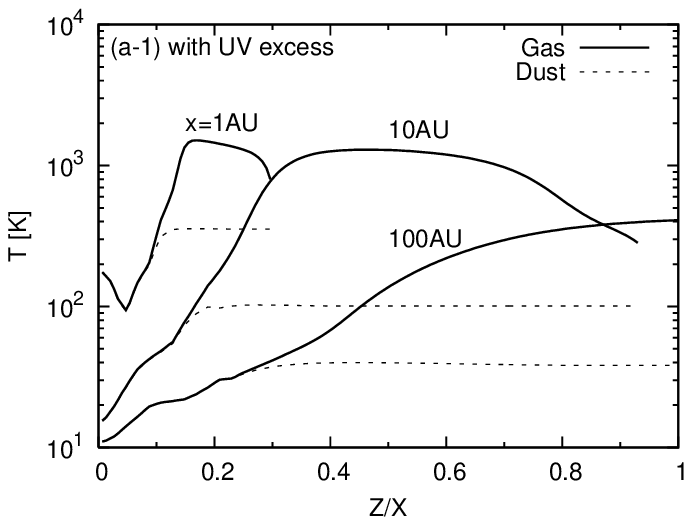}
\includegraphics[width=8cm]{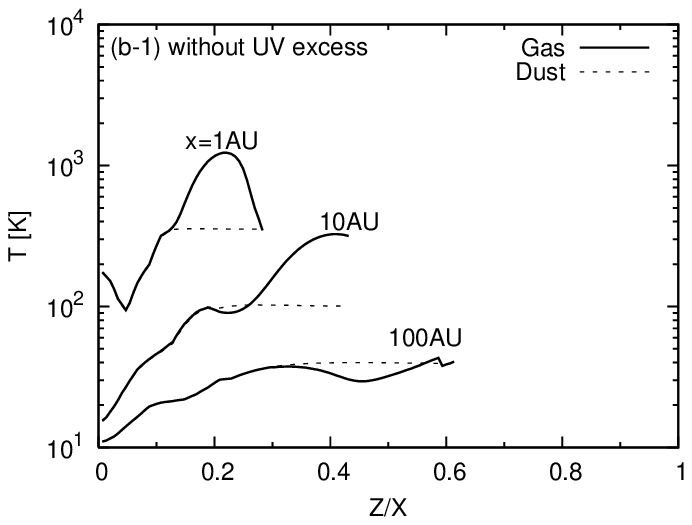}
\includegraphics[width=8cm]{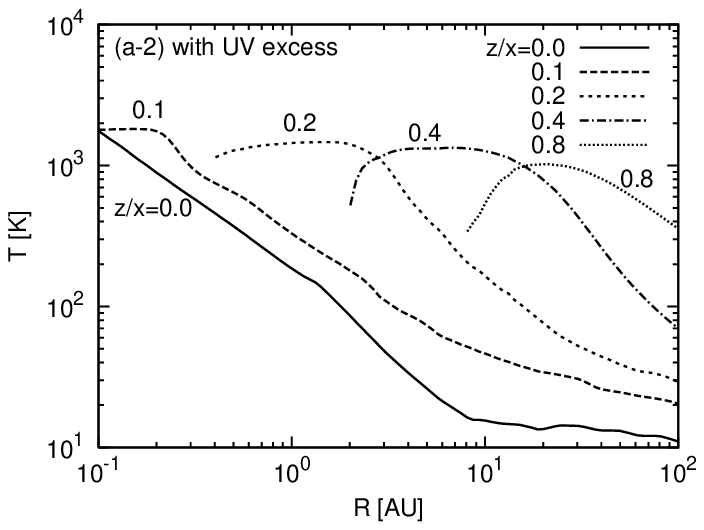}
\includegraphics[width=8cm]{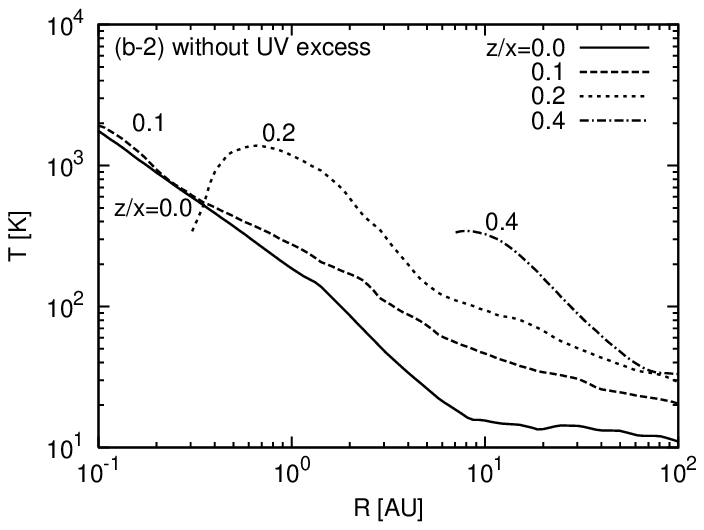}
\caption{The vertical temperature profiles of gas (solid lines) and dust
 (dotted lines) at the disk radii of $x=1, 10$ and 100AU {(a-1) (b-1),
 and the radial temperature profiles of gas at $z/x=0.0, 0.1, 0.2, 0.4$,
 and 0.8 (a-2) (b-2)} for the models (a) with and (b) without UV excess. 
 In the model (a) the gas temperature at the disk
 surface can reach around 1,000 K due to the grain photoelectric
 heating, while it is almost the same as the dust temperature near the
 midplane.} 
\label{f2.2}
\end{figure*}
%
\begin{figure*}
\centering
\includegraphics[width=8cm]{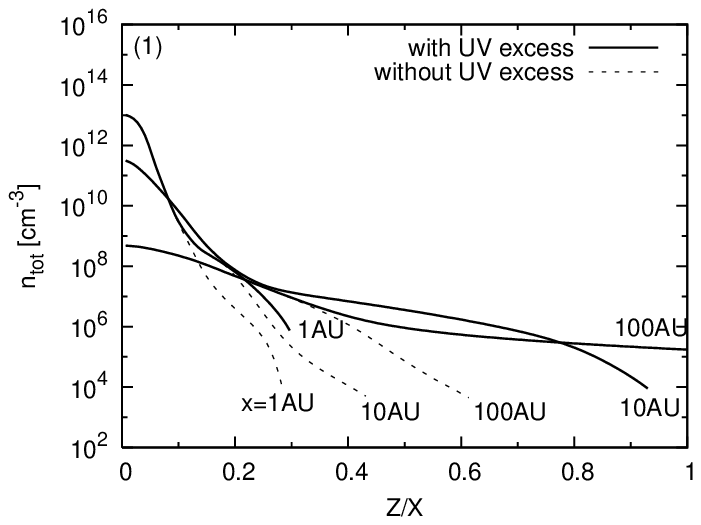}
\includegraphics[width=8cm]{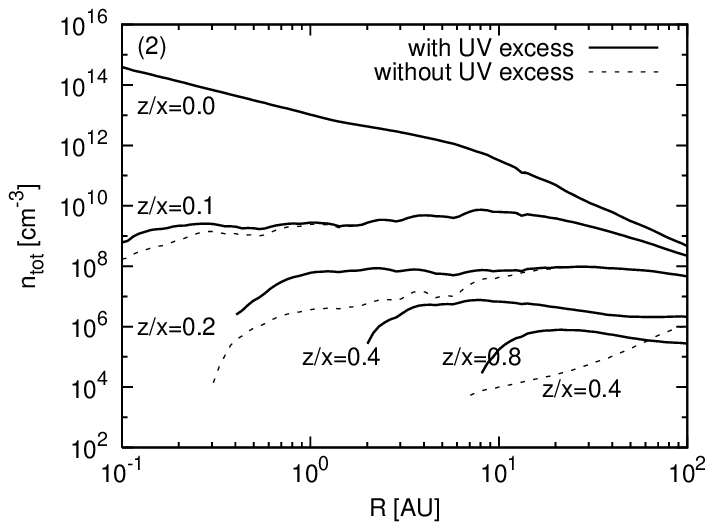}
\caption{The vertical gas density profiles at the disk radii of 
 $x=1, 10$ and 100 AU {(1), and the radial gas density profiles
 at $z/x=0.0, 0.1, 0.2, 0.4$, and 0.8 (2)} for the models with
 (solid lines) and without (dotted lines) UV excess. 
 The UV irradiation heats the gas and makes the disk expand vertically,
 which results in higher density at the disk surface.}
\label{f2.3}
\end{figure*}

Now, it is known observationally that many classical T Tauri stars
have excess continuum radiation in the ultraviolet
region, compared to the main-sequence stars of similar effective
temperature (e.g., Herbig \& Goodrich 1986; Herbst et al. 1994;
Valenti, Johns-Krull, \& Linsky 2000). The UV excess radiation is
considered to result from the shock on the stellar surface which is
caused by a magnetospherical accretion flow from the accretion disk onto
the star (e.g., Calvet \& Gullbring 1998; Ostriker \& Shu 1995). 
In order to examine the effects of this UV excess radiation on the disk
structure and the H$_2$ line emission from the disk, we treat two
kinds of models for radiation from the central star, $F_{\nu,
{\rm star}}$: models with and without UV excess. 
In the model with UV excess, $F_{\nu, {\rm star}}$ consists of
three components: black body emission at the star's effective temperature,
$T_*$, an optically thin hydrogenic thermal bremsstrahlung emission at a
higher temperature, and Ly $\alpha$ line emission, based on observations
towards TW Hydrae (see Appendix C). In the model without UV excess, the
stellar radiation is assumed to be black body emission only. 

For the interstellar radiation field, $F_{\nu, {\rm ISRF}}$, we adopt 
the Draine (1978) field for the wavelength range of $912{\rm \AA} < \lambda <
2000$\AA\ and the field given by van Dishoeck \& Black (1982) for $\lambda
> 2000$\AA.

\begin{figure*}
\centering
\includegraphics[width=8cm]{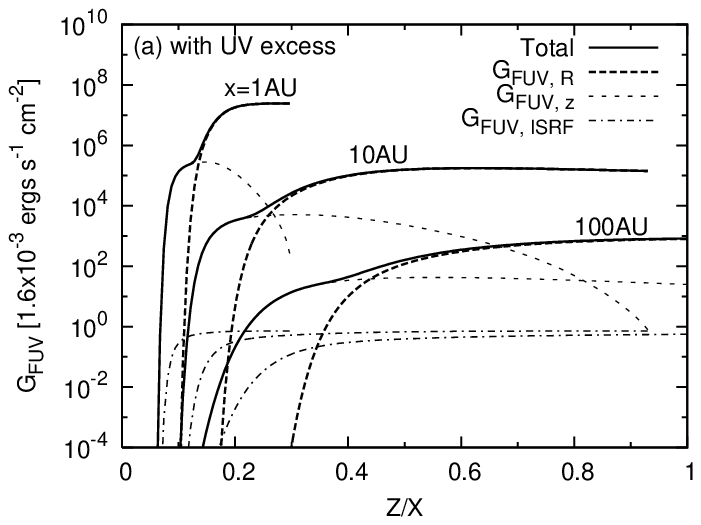}
\includegraphics[width=8cm]{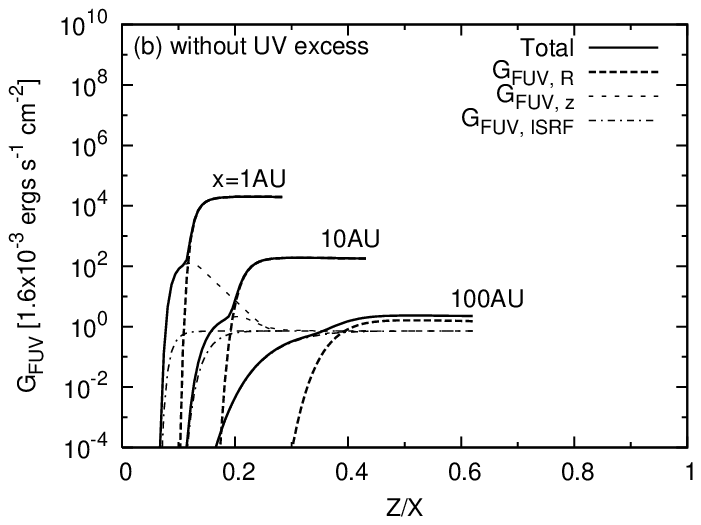}
\caption{{The integrated FUV radiation fields (6 eV $<h\nu<$ 13 eV)}
 as a function of the vertical height 
 of the disk at the radii of $x=1, 10$ and 100AU for the models (a) with
 and (b) without UV excess. The solid, dashed and dotted lines show
 the total, radial, and vertical radiation fields, respectively. The
 contributions of 
 the interstellar radiation fields are plotted in the dot-dashed lines.}
\label{f2.4}
\end{figure*}
%
\begin{figure*}
\centering
\includegraphics[width=8cm]{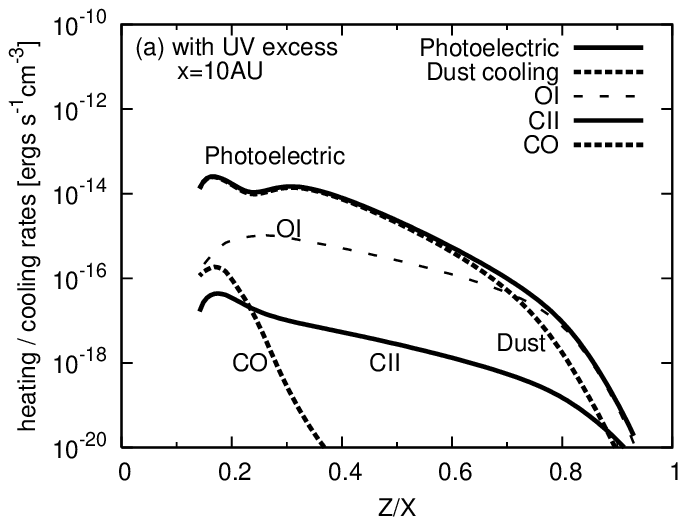}
\includegraphics[width=8cm]{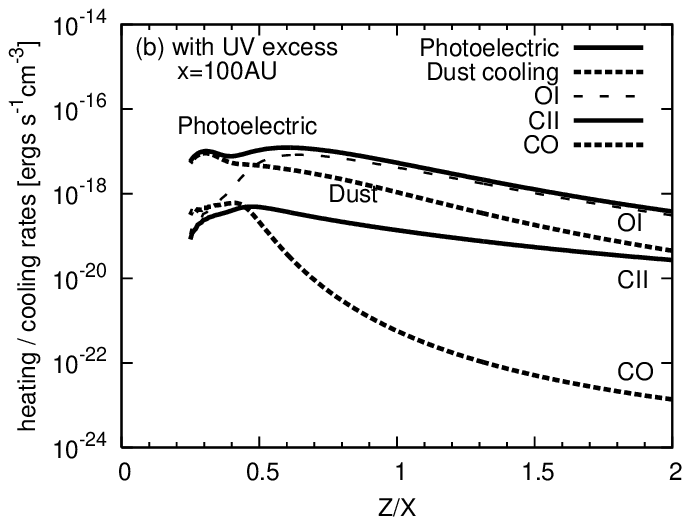}
\caption{The vertical cooling and heating rates at the disk radii of (a)
 $x=10$ AU and (b) 100 AU for the models with UV excess. The grain
 photoelectric heating dominates the heating process,
 while the gas-grain collisions and the radiative cooling dominate the
 cooling process in the dense and less dense regions, respectively.}
\label{f2.5}
\end{figure*}

\subsection{Dust model}
The physical and chemical structure of protoplanetary disks is affected
by the dust model in various ways: first, it affects the
radiation field because dust grains are the dominant opacity
source in protoplanetary disks. Thus, the dust temperature profile is
influenced by the dust model through the absorption ($\kappa_{\nu}$) and
extinction ($\chi_{\nu}$) coefficients (Sect. 2.1).
Also, it affects the UV radiation field (through the extinction
coefficient $\chi_{\nu}$ and albedo $\omega_{\nu}$; Sect. 2.2), 
that is, the photodissociation
and photoionization processes (see Appendix B and Sect. 3). The gas
temperature profile is also affected because grain photoelectric 
heating induced by
UV photons is the dominant source of gas heating at the disk surface
(see Appendix A.1 and Sect. 2.4) and by grain-gas collisions (see Appendix 
A.2).
Second, the total surface area of the dust grains
affect the molecular hydrogen abundance through
the H$_2$ formation rate on grain surfaces (Sect. 3).

In this paper we assume that the dust particles consist of silicate,
carboneous grains, and water ice, and have size distributions obtained
by Weingartner \& Draine (2001a; hereafter WD01a) which reproduce the
observational extinction curve of dense clouds (see Appendix D for
details). Also, the dust and gas are assumed to be well-mixed.

\subsection{Resulting temperature and density profiles}

 We obtain self-consistent density and temperature
distributions of a protoplanetary disk by solving the equations for 
vertical hydrostatic equilibrium and local
thermal balance between heating and cooling of gas iteratively together
with the local radiative equilibrium equations on dust grains (see
Sect. 2.1). Figure \ref{f2.1}a shows the
contour plots of the resulting temperature {(a-1)} and
density {(a-2)} profiles in the $x$-$z$ plane. The contour
levels are $T=30,100,300$, and $1000$K for the temperature, and
$\rho=10^{-20},10^{-18},10^{-16},10^{-14}$, and $10^{-12}$g cm$^{-3}$
for the density. In order to see the effects of the UV excess radiation,
we also calculate for comparison the structure of the disk which is
irradiated by stellar radiation without UV excess (Fig.~\ref{f2.1}b;
see Sect. 2.2). 

\begin{figure*}
\centering
\includegraphics[width=8cm]{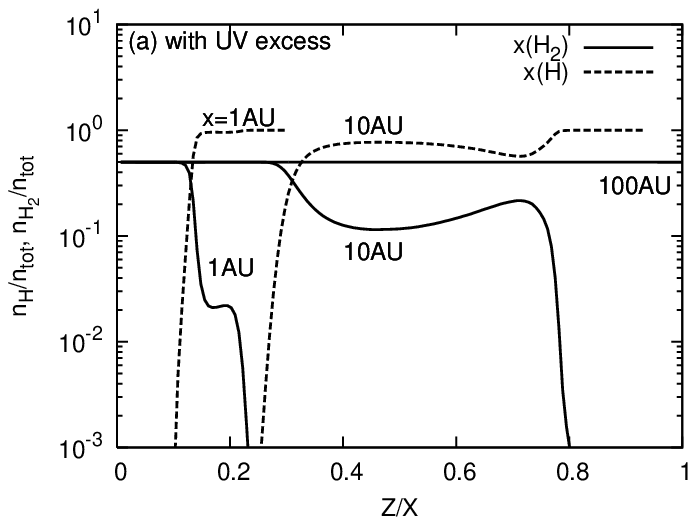}
\includegraphics[width=8cm]{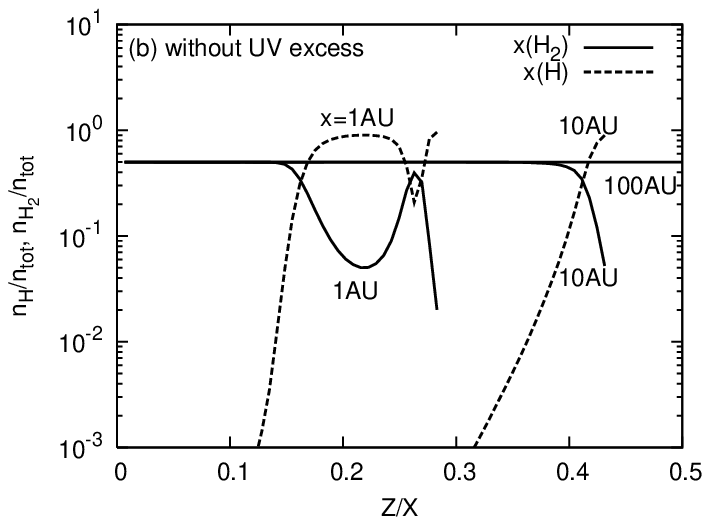}
\caption{The vertical molecular (solid lines) and atomic (dashed lines)
 hydrogen abundances at the disk radii of $x=1, 10$ and 100 AU for the
 models (a) with and (b) without UV excess radiation.
 The molecular hydrogen is photodissociated at the very disk surface and
 destroyed by atomic oxygen where the gas temperature is $\sim 1,000$ K.}
\label{f3.1}
\end{figure*}

Figure \ref{f2.2} shows the corresponding gas (solid lines) and dust
(dotted lines) temperature profiles in the vertical direction at
the radii of $x=1, 10$ and 100 AU {(a-1) (b-1)}. The horizontal axis
is the vertical height divided by each radius. {The gas temperature
profiles as a function of the distance from the central star,
$R=(x^2+z^2)^{1/2}$, are also plotted in (a-2) and (b-2) for $z/x=0.0,
0.1, 0.2, 0.4$, and 0.8.} The figures show that in the
model with UV excess radiation the gas temperature at the disk
surface is much higher than the dust temperature and reaches around
1,000K even at the radius of 10AU. This is due to the grain
photoelectric heating induced by the strong far ultraviolet (FUV)
radiation from the central star. On the other hand, near the midplane of 
the disk and at the outer disk in the model without UV excess, where
the UV radiation field is not strong enough to heat up the gas, the gas
temperature is close the dust temperature due to the collisions between
the gas and dust particles (see below). We calculate the gas temperature
only in the upper region of the disk where the difference between the
gas and dust temperatures are more than 1\% ($|T-T_d|/T>10^{-2}$),
considering that the gas temperature is almost the same as the dust
temperature near the midplane. The gas temperature in ionized region is
assumed to be $T=10^4$ K (see Appendix B.1).
Now, the corresponding vertical {(1) and radial (2)} density
profiles of the disk in the models with (solid lines) and without
(dotted lines) UV excess radiation are plotted in Fig.~\ref{f2.3}.
The strong UV radiation field at the disk surface heats the gas hot
enough to make the disk expand in the vertical direction. Therefore, the
higher density at the disk surface in the model with UV excess
radiation is caused by the stronger UV radiation fields.

The {integrated} FUV radiation fields {(6 eV $<h\nu<$ 13 eV)} in
the disk for the models with and without
UV excess radiation are shown in Fig.~\ref{f2.4}. 
{The radial (dashed lines), $G_{\rm FUV, R}$, and vertical (dotted
lines), $G_{\rm FUV, z}$, fields are calculated by integrating the
specific radiation field $F_{\nu, R}$ (Eq. [\ref{eq.2-6}]) and $F_{\nu,
z}$ (Eq. [\ref{eq.2-7}]), respectively, in the FUV region (6 eV $<h\nu<$
13 eV). The total radiation fields (solid lines) are summations of
the radial and vertical fields, $G_{\rm FUV}=G_{\rm FUV, R}+G_{\rm FUV,
z}$.}
The figures show that the direct irradiation from the central star
dominates the radiation field in the upper layer, while the scattered
field is superior in the lower layer (see also van Zadelhoff et
al. 2003; Bergin et al. 2003). The contribution of the interstellar
radiation field (dot-dashed lines), $G_{\rm FUV, ISRF}=\int
F_{\nu, {\rm ISRF}}\exp[-\tau_{\nu, z}(z_{\infty})]d\nu$, is also
plotted in this 
figure, which shows that it is not dominant except in the lower layer of
the outer region ($x\sim 100$ AU) in the model without UV excess
radiation. 

The vertical profiles of the heating and cooling rates at the radii of
(a) 10AU and (b) 100AU for the model with UV excess radiation are
plotted in Fig.~\ref{f2.5}. We can see from the figure that the dominant
heating source is grain photoelectric heating in the region where we
calculate the gas temperature ($|T-T_d|/T>10^{-2}$).
Meanwhile, the energy transfer from gas to dust particles via
collisions is the dominant cooling source at the inner region and near
the midplane of the disk where the matter is dense enough, while 
radiative cooling via line transition is dominant in less dense region.

\section{Abundance and excitation of H$_2$}

\begin{figure*}
\centering
\includegraphics[width=8cm]{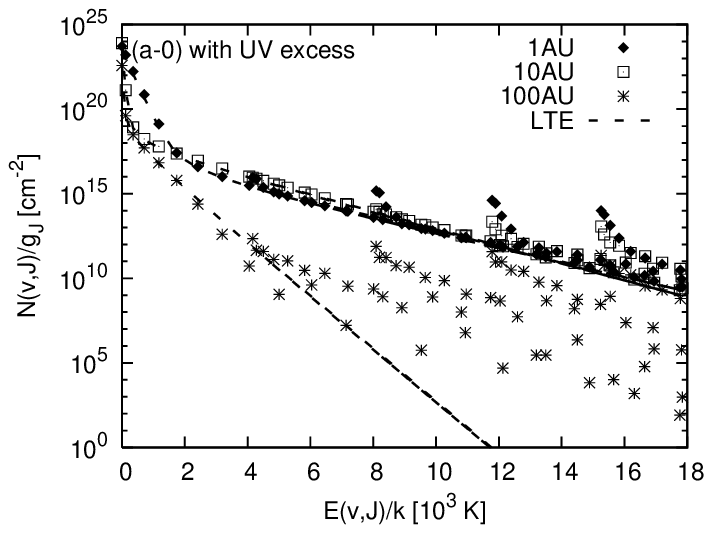}
\includegraphics[width=8cm]{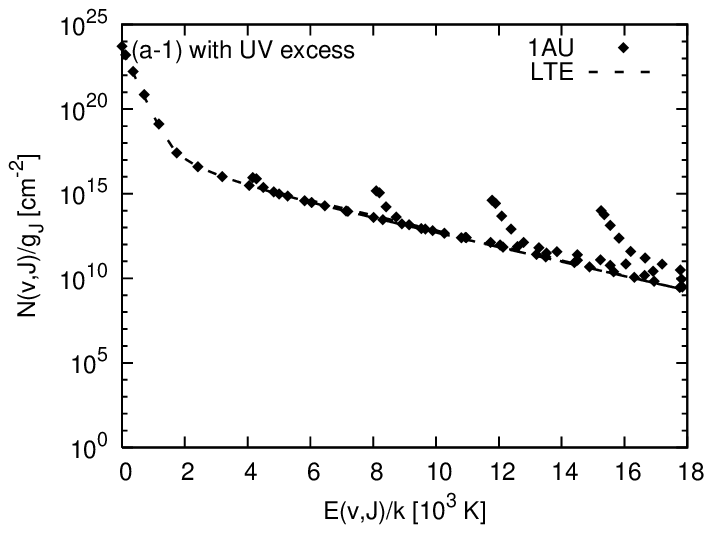}
\includegraphics[width=8cm]{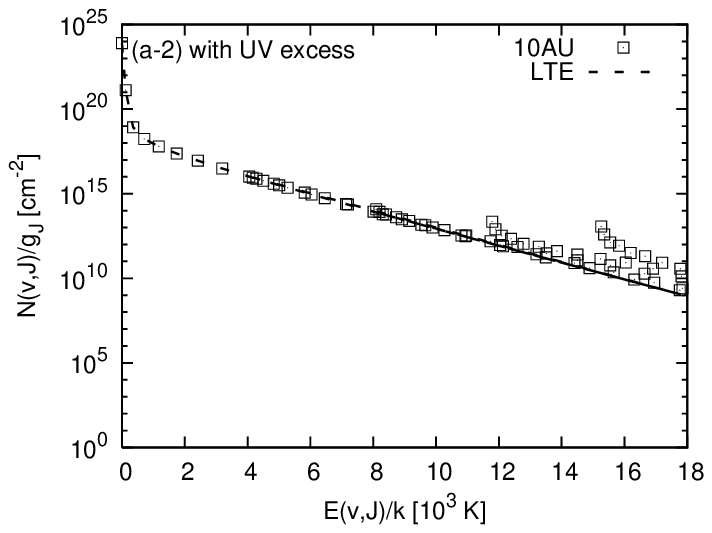}
\includegraphics[width=8cm]{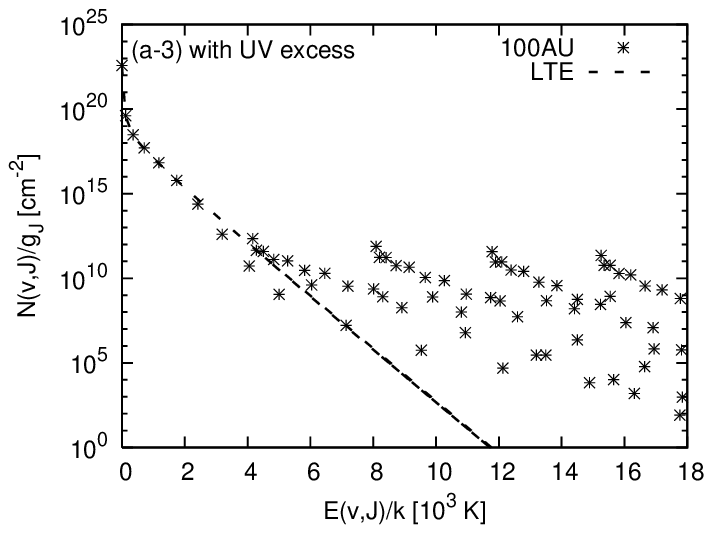}
\includegraphics[width=8cm]{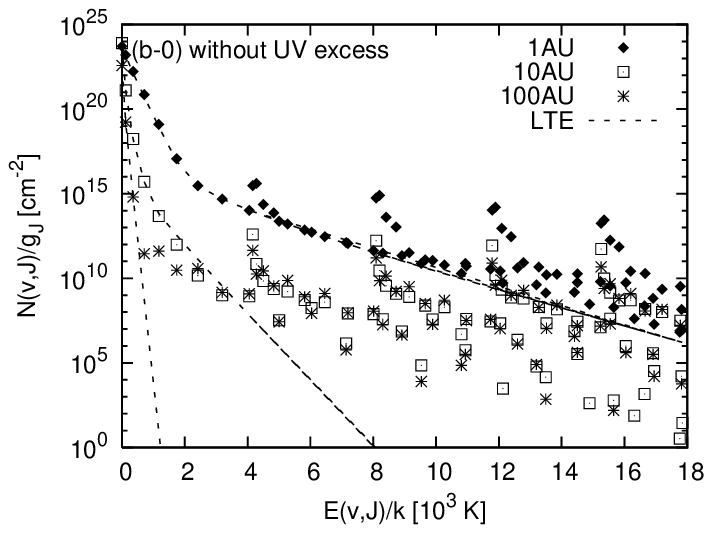}
\includegraphics[width=8cm]{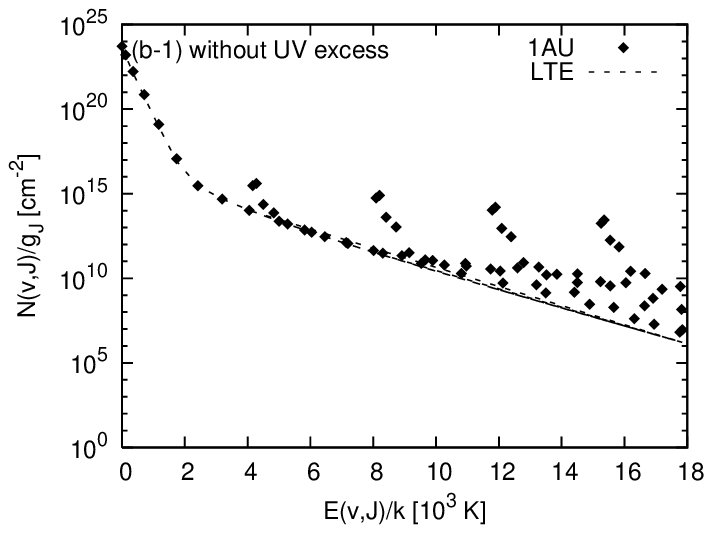}
\includegraphics[width=8cm]{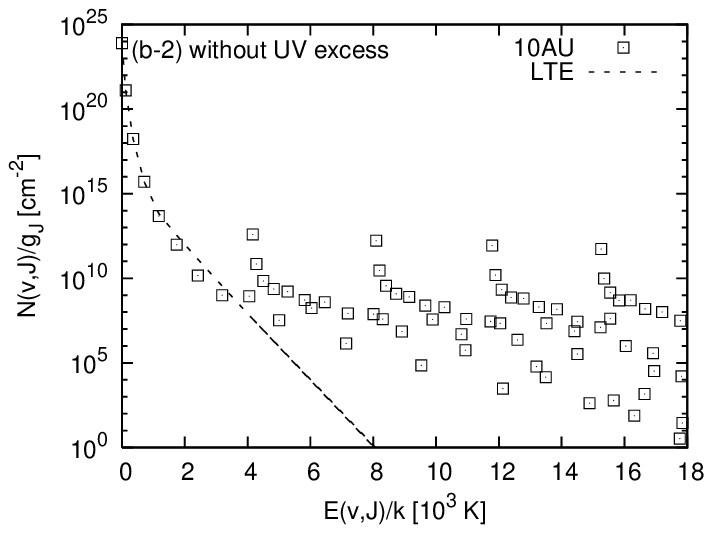}
\includegraphics[width=8cm]{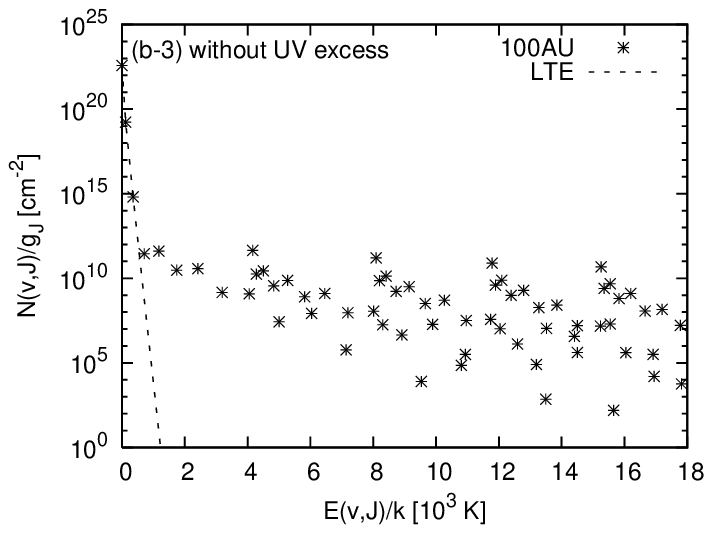}
\caption{The level populations of molecular hydrogen at the disk radii
 of $x=1$ AU (filled diamonds), 10AU (open squares) and 100AU
 (asterisks) for the models (a) with and (b) without UV excess
 radiation. The LTE distributions are plotted in dashed lines.
 If the central star has a UV excess, the level populations are in
 LTE at the inner disk, and the populations of the upper levels become
 large due to the high temperature.}
\label{f3.2}
\end{figure*}

\subsection{The model}

The abundance and the level populations of the $X^1\Sigma_g^-$ electronic
state of molecular hydrogen in a statistical equilibrium state are
calculated based on Wagenblast \& Hartquist (1988) as
\begin{displaymath}
n_l({\rm H}_2)\left[\sum_{m\ne l} \biggl(A_{lm}+\beta_{lm}+\sum_s n_sC_{lm}^s\biggr)+R_{{\rm diss},l}\right]
\end{displaymath}
\begin{displaymath}
\hspace*{4.5cm} +k_{{\rm O}+{\rm H}_2}n({\rm O})n_l({\rm H}_2)=
\end{displaymath}
\begin{equation}
\sum_{m\ne l}n_m({\rm H}_2)\biggl(A_{ml}+\beta_{ml}+\sum_s n_sC_{ml}^s\biggr)+n({\rm H})R_{{\rm form},l} \label{eq.3-1}
\end{equation}
where $A_{lm}$ is the Einstein $A$-coefficient for spontaneous
emission from level $l$ to level $m$ ($A_{lm}=0$ if the energy of
level $l$, $E_l$, is smaller than $E_m$)
and $C_{lm}^s$ is the collisional transition rate
with collision partner $s$. The collisional rate coefficients are taken
from Smith et al. (1982) and Black \& Dalgarno (1977) for
collisions with H$^+$,  Tin\'e et al. (1997; see also references therein)
for the collisions with H$_2$, and Lepp et al. (1995; see also Tin\'e et
al. 1997) for the collisions with H
(http://www.physics.unlv.edu/astrophysics/). 

The symbol $\beta_{lm}$ in Eq. (\ref{eq.3-1}) represents the
effective rate for transition $l\rightarrow m$ via ultraviolet pumping
followed by radiative cascades and $R_{{\rm diss},l}$ is the
photodissociation rates of hydrogen molecules in level $l$.
The UV radiation fields are calculated as discussed in Sect. 2.2.
An approximate function given by Federman et al. (1979) is used for the
H$_2$ self-shielding of the UV radiation fields.
A purely thermal Doppler widths of $v_d=(2kT/m_{\mu})^{0.5}$ is adopted
in this function. {Molecular hydrogen excitation due to X-ray
induced electron impact should be taken into account in future (e.g.,
Bergin et al. 2004).}

The symbol $R_{{\rm form},l}$ represents the effective formation rate
of H$_2$ in level $l$ on grain surfaces. The molecular hydrogen which
leaves 
the grain surfaces is assumed to originate in the levels $(v=7,J=1)$
and $(v=7,J=0)$ with the ratio of 9:1 and cascade into the level $l$,
following Duley \& Williams (1986) (cf. Takahashi \& Uehara 2001 for new
formation pumping model). The total formation rate of $\sum_l
R_{{\rm form},l}=7.5\times 10^{-18}T^{0.5}\epsilon_{{\rm H}_2}n_{\rm 
tot}n(H)$ 
cm$^{-3}$ s$^{-1}$ is adopted here. The symbol $\epsilon_{{\rm H}_2}$
represents the recombination efficiency of atomic hydrogen on dust
grains which is estimated by Cazux \& Tielens (2002, 2004) based on
laboratory experiments (e.g., Pirronello et al. 1999; Zecho et al. 2002).

The endothermic reaction O + H$_2 \rightarrow$ OH + H, which dominates the 
destruction of molecular hydrogen in
high temperature regions (e.g., Storzer \& Hollenbach 1998), is also taken into
account. The reaction
rate coefficient of $k_{{\rm O}+{\rm H}_2}=3.14\times
10^{-13}(T/300K)\exp(-3150K/T)$ cm$^3$ s$^{-1}$ is adopted from the 
UMIST RATE99 database (Le Teuff et al. 2000). 

The total number density of hydrogen nuclei at each point in the disk is
required to satisfy the condition,
\begin{equation}
n_{\rm tot}=n({\rm H})+2n({\rm H}_2)+n({\rm H}^+)+2n({\rm H}_2^+)+3n({\rm H}_3^+),
\end{equation}
where $n(s)$ represents the number density of species $s$. A simple
chemical equilibrium scheme given in Wagenblast \& Hartquist (1988) is
used to obtain the number densities of the related species.

\subsection{Results}

The resulting molecular (solid lines) and atomic (dashed lines) hydrogen
abundances in the vertical direction at the radii of $x=1, 10$ and 100 AU 
are plotted in Fig.~\ref{f3.1}. Molecular hydrogen is
photodissociated at the very surface of the disk, while it is destroyed by
atomic oxygen where the gas temperature is $\sim 1,000$ K.

Figure \ref{f3.2} shows the resulting level populations of molecular
hydrogen. The filled 
diamonds, open squares, and asterisks show the column densities of molecular
hydrogen in each ro-vibrational level as a function of the level energy
at each radius of 1, 10 and 100 AU, respectively. The column densities are
calculated by integrating the number density of molecular hydrogen in
each level along the vertical direction at each radius of the disk.
The level populations in local thermodynamic
equilibrium (LTE) at each radius are shown in the dashed lines. Also, 
we present the level populations in the model without UV excess in
Fig.~\ref{f3.2}b for comparison. 
{Figures labeled (1), (2), and (3) show the level populations at
the radius of 1AU, 10AU, and 100AU, respectively, and they are plotted
together in figure (0) for comparison.}
We can see from the figures that 
if the central star has a UV excess, the gas becomes hot enough, 
as we have seen in the previous section, that the collisional processes
are very efficient and the level populations are in 
LTE except for those of the very high levels. And as a result of the
high temperature, the populations of the upper levels become very large.
On the other hand, in the outer disk in the model without UV 
excess where the gas is cold enough, the level populations are 
not in LTE but are affected by the UV pumping process.
Also we can see that in this model, the populations of the upper 
levels are not as large as those in the model with UV excess.

\section{Molecular hydrogen emission}

Making use of the level populations we obtained in Sect. 3, we calculate
the molecular hydrogen emission from the disk. The intensity of each
line integrated over the frequency, $I_{ul}$, is obtained
by solving the radiative transfer equation in the vertical direction of the
disk, assuming that an observer faces the disk, 
\begin{equation}
\dfrac{dI_{ul}}{dz}=-\chi_{ul}(I_{ul}-S_{ul}), \label{eq.4-1}
\end{equation}
where the subscripts $ul$ means the transition from the upper to the
lower levels. The source function, $S_{ul}$, and the total extinction
coefficient, $\chi_{ul}$, are given as
\begin{equation}
S_{ul}=\dfrac{1}{\chi_{ul}}n_uA_{ul}\Phi_{ul}\dfrac{h\nu_{ul}}{4\pi} \label{eq.4-2}
\end{equation}
and
\begin{equation}
\chi_{ul}=\rho\chi_{\nu_{ul}}+(n_lB_{lu}-n_uB_{ul})\Phi_{ul}\dfrac{h\nu_{ul}}{4\pi}, \label{eq.4-3}
\end{equation}
where the symbols $A_{ul}$ and $B_{ul}$ are the Einstein coefficients
for the transition $u\rightarrow l$, and $n_u$ and $n_l$ are the number
densities of the upper and lower levels, respectively. The energy
difference between the levels $u$ and $l$ corresponds to $h\nu_{ul}$. 
The symbol $\Phi_{ul}$ is the line profile function, $\rho$ the gas
density, and $\chi_{\nu_{ul}}$ the dust opacity at the frequency
$\nu_{ul}$.

\begin{figure*}
\centering
\includegraphics[width=8cm]{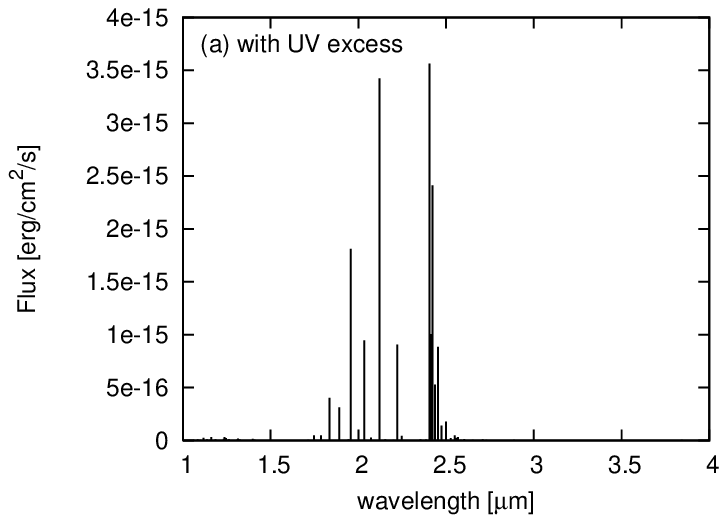}
\includegraphics[width=8cm]{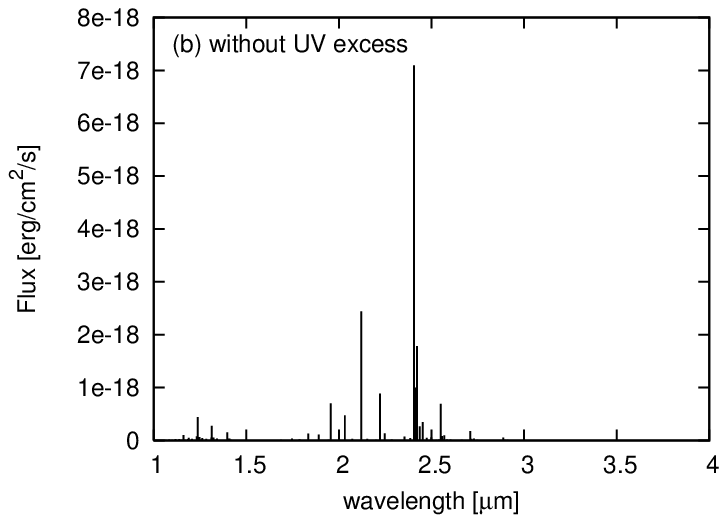}
\caption{The near-infrared ($1\mu m<\lambda <4\mu m$) line spectra of
 molecular hydrogen from the disk in the model
 (a) with and (b) without UV excess radiation of the central star.}
\label{f4.1}
\end{figure*}
%
\begin{figure*}
\centering
\includegraphics[width=8cm]{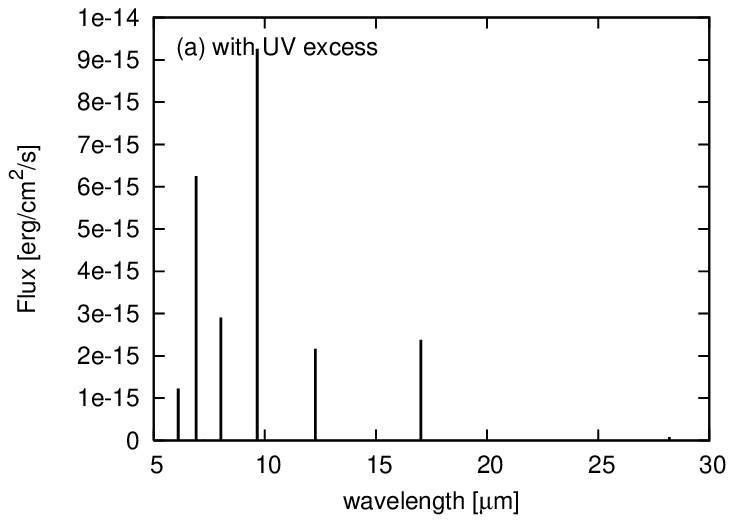}
\includegraphics[width=8cm]{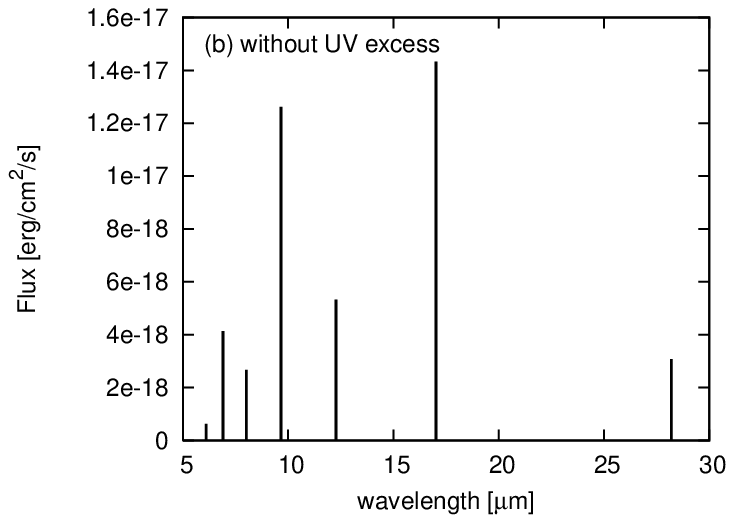}
\caption{Same as Fig.~\ref{f4.1}, but for the mid-infrared ($5\mu m<\lambda <30\mu m$) wavelength band.}
\label{f4.2}
\end{figure*}

The observable line flux is obtained by integrating Eq. (\ref{eq.4-1})
in the vertical direction and summing up the integrals in the radial
direction of the disk as,
\begin{equation}
F_{ul}=\dfrac{1}{4\pi d^2}\int_{x_{\rm in}}^{x_{\rm out}}2\pi xdx\int_{-z_{\infty}}^{z_{\infty}}\tilde{\eta}_{ul}(x,z)dz, \label{eq.4-4}
\end{equation}
{where $\tilde{\eta}_{ul}(x,z)$ is the emissivity of the transition
line at $(x,z)$ times the effect of absorption in the upper disk layer,}
\begin{equation}
\tilde{\eta}_{ul}(x,z)=n_u(x,z)A_{ul}\dfrac{h\nu_{ul}}{4\pi}\exp(-\tau_{ul}(x,z)), \label{eq.4-5}
\end{equation}
and $\tau_{ul}(x,z)$ is the optical depth from $z$ to the disk surface
$z_{\infty}$ at the frequency $\nu_{ul}$,
\begin{equation}
\tau_{ul}(x,z)=\int_z^{z_{\infty}}\chi_{ul}(x,z')dz'. \label{eq.4-6}
\end{equation}
Here, we use the distance to an object of $d=56$ pc for calculating the
intensity in order to compare it with the observations towards TW Hya.

\subsection{The infrared spectra}

The resulting line spectra in the near- and mid-infrared wavelength
bands are plotted in Figs.~\ref{f4.1}a and \ref{f4.2}a, respectively.
The line spectra for the model without UV excess are
presented in Figs.~\ref{f4.1}b and \ref{f4.2}b for comparison. We
can see from the figures that if the central star has a UV excess, the
line spectra become stronger by about a few orders of magnitude because the
strong UV irradiation by the central star heats the gas hot enough and
makes the populations of the upper levels large, as we have seen in
Sect. 3.2. 

\begin{table*}
 \caption[]{The observed and calculated infrared line flux of molecular
 hydrogen ($F_{\rm line}$), the calculated continuum radiation flux
 ($F_{\rm cont}$) [erg/s/cm$^2$], and the line-to-continuum flux ratio
 ($F_{\rm line}/F_{\rm cont}$).} \label{T4.1}
 $$ 
 \begin{array}{cc|ccc|cc} 
  \hline 
   \lambda  & {\rm Line} & \multicolumn{3}{|c|}{F_{\rm line}} & F_{\rm cont.} & F_{\rm line}^{\mathrm{d}}/F_{\rm cont.} \\
   & & {\rm Model\ A} & {\rm Model\ B} & {\rm Observation} & & \\
  (\mu m)  &  & (\times 10^{-15}) & (\times 10^{-18}) & (\times 10^{-15}) & (\times 10^{-9}) & (\times 10^{-6}) \\
   \hline 
  1.16 &  2-0\ S(1) & 0.03 &  0.10 &  &  5.19 &  0.01 \\
  1.19 &  2-0\ S(0) & 0.01 &  0.05 &  &  5.09 &  0.00 \\
  1.24 &  2-0\ Q(1) & 0.03 &  0.44 &  &  4.91 &  0.01 \\
  1.24 &  2-0\ Q(2) & 0.01 &  0.05 &  &  4.90 &  0.00 \\
  1.25 &  2-0\ Q(3) & 0.02 &  0.06 &  &  4.87 &  0.00 \\
  1.83 &  1-0\ S(5) & 0.40 &  0.13 &  &  2.87 &  0.14 \\
  1.89 &  1-0\ S(4) & 0.31 &  0.11 &  &  2.73 &  0.11 \\
  1.96 &  1-0\ S(3) & 1.81 &  0.70 &  &  2.57 &  0.71 \\
  2.03 &  1-0\ S(2) & 0.95 &  0.47 &  &  2.40 &  0.39 \\
  2.12 &  1-0\ S(1) & 3.43 &  2.44 & 1.0^{\mathrm{a}} &  2.22 &  1.55 \\
  2.22 &  1-0\ S(0) & 0.91 &  0.89 &  &  2.03 &  0.45 \\
  2.25 &  2-1\ S(1) & 0.04 &  0.13 &  &  1.99 &  0.02 \\
  2.35 &  2-1\ S(0) & 0.01 &  0.07 &  &  1.81 &  0.01 \\
  2.41 &  1-0\ Q(1) & 3.56 &  7.09 &  &  1.74 &  2.05 \\
  2.41 &  1-0\ Q(2) & 1.00 &  1.00 &  &  1.73 &  0.58 \\
  2.42 &  1-0\ Q(3) & 2.41 &  1.78 &  &  1.71 &  1.41 \\
  2.44 &  1-0\ Q(4) & 0.53 &  0.27 &  &  1.69 &  0.31 \\
  2.45 &  1-0\ Q(5) & 0.88 &  0.34 &  &  1.67 &  0.53 \\
  2.47 &  1-0\ Q(6) & 0.14 &  0.05 &  &  1.64 &  0.09 \\
  2.50 &  1-0\ Q(7) & 0.18 &  0.06 &  &  1.61 &  0.11 \\
  2.55 &  2-1\ Q(1) & 0.05 &  0.69 &  &  1.55 &  0.03 \\
  2.56 &  2-1\ Q(2) & 0.01 &  0.08 &  &  1.54 &  0.01 \\
  2.57 &  2-1\ Q(3) & 0.03 &  0.10 &  &  1.52 &  0.02 \\ \hline
  6.11 &  0-0\ S(6) & 1.22 &  0.62 &  &  0.25 &  4.98 \\
  6.91 &  0-0\ S(5) & 6.25 &  4.14 &  &  0.19 & 32.06 \\
  8.02 &  0-0\ S(4) & 2.90 &  2.67 &  &  0.18 & 16.10 \\
  9.66 &  0-0\ S(3) & 9.26 & 12.62 &  &  0.24 & 38.37 \\
 12.27 &  0-0\ S(2) & 2.17 &  5.33 &  <30^{\mathrm{c}} &  0.15 & 14.28 \\
 17.02 &  0-0\ S(1) & 2.38 & 14.33 & 28-81^{\mathrm{b}}, <39^{\mathrm{c}} &  0.13 & 18.62 \\
 28.20 &  0-0\ S(0) & 0.08 &  3.08 & 25-57^{\mathrm{b}} &  0.09 &  0.94 \\
   \hline 
 \end{array}
 $$ 
 \begin{list}{}{}
  \item[$^{\mathrm{a}}$] Observation by Bary et al. (2003).
  \item[$^{\mathrm{b}}$] Observations by Thi et al. (2001b).
  \item[$^{\mathrm{c}}$] Observations by Richter et al. (2002).
  \item[$^{\mathrm{d}}$] $F_{\rm line}$ in Model A.
 \end{list}
\end{table*}

Some of the strongest line fluxes ($F_{\rm line}$) we calculate for the
models with (Model A) and without (Model B) UV excess radiation are
listed in 
Table~\ref{T4.1}. The intensity ratios of the $2.12 \mu m\ v=1\rightarrow
0\ S(1)$ line to the $2.25 \mu m\ v=2\rightarrow 1\ S(1)$ line are $\sim
80$ and $\sim 18$ for Model A and B,
respectively. The corresponding excitation temperatures are $\sim 900$ K
and $\sim 1300$ K, which are comparable to the kinetic temperature
(cf. Fig.~\ref{f2.2}) as the level populations are in LTE (see
Fig.~\ref{f3.2}). The excitation temperature is higher in the model
without UV excess than in the model with UV excess because 
the emission comes mainly from the inner disk in the former case while
it comes mostly from the outer cooler disk in the latter case as we will
see below. The line ratio of the $2.12 \mu m\ v=1\rightarrow 0\ S(1)$
line to the $2.41 \mu m\ v=1\rightarrow 0\ Q(1)$, two of the strongest
lines in Fig.~\ref{f4.1}, is larger for the model with UV excess for the
same reason.
On the other hand, the pure rotational transition lines in
Fig.~\ref{f4.2} are emitted from all over the disk for both models since
they trace the gas in cooler regions than the vib-rotational transition
lines in Fig.~\ref{f4.1} (see below). Therefore, the excitation
temperature derived from the pure rotational transition lines for 
the model with UV excess is higher (that is, the weaker line flux
for the lower level transitions) than that derived from the lines for
the model without UV excess because the UV irradiation from the
central star heats the gas in the disk more in the former model.

The observational intensity of the ro-vibrational line,
$v=1\rightarrow 0\ S(1)$, towards TW Hya by Bary et al. (2003) is also
presented in Table~\ref{T4.1}, and is comparable to our calculated 
flux in the model with UV excess. In addition, the observations
of mid-infrared line spectra of pure rotational transitions, $S(0)$ and
$S(1)$, by ISO (Thi et al. 2001b) and the upper limit of the line fluxes
of $S(1)$ and $S(2)$ under the assumption of FWHM of 30 km s$^{-1}$
observed by TEXES at the NASA IRTF (Richter et al. 2002) are presented
in Table~\ref{T4.1}. These are observed towards classical T Tauri stars
in the Taurus-Auriga cloud complex ($d=140$pc). We can see from the
table that the calculated line fluxes are weaker than the results of the
ISO observations, and consistent with the ground-based observations by
TEXES.
It may be that the spatial and wavelength resolutions of the ISO
observations were not high enough so that the observed line emission comes
not from the circumstellar disk but from shocked regions in the
associated outflows.
The last two columns of Table~\ref{T4.1} represent the flux of the
calculated dust continuum emission from the disk plus the radiation from
the central star ($F_{\rm cont.}$) and the flux ratio of the molecular
hydrogen line in Model A to the continuum emission ($F_{\rm
line}/F_{\rm cont.}$).
The dust continuum emission is calculated by solving the radiative
transfer equation (Eq. [\ref{eq.4-1}]), adopting
$S_{ul}=B_{\nu_{ul}}(T_d)$ and $\chi_{ul}=\rho\chi_{\nu_{ul}}$ for
the source term and the opacity, respectively. The central star is
assumed to emit blackbody radiation with an effective temperature of
$T_*=4000$K and a radius of $R_*=2R_{\odot}$.

\begin{figure}
\centering
\resizebox{\hsize}{!}{\includegraphics{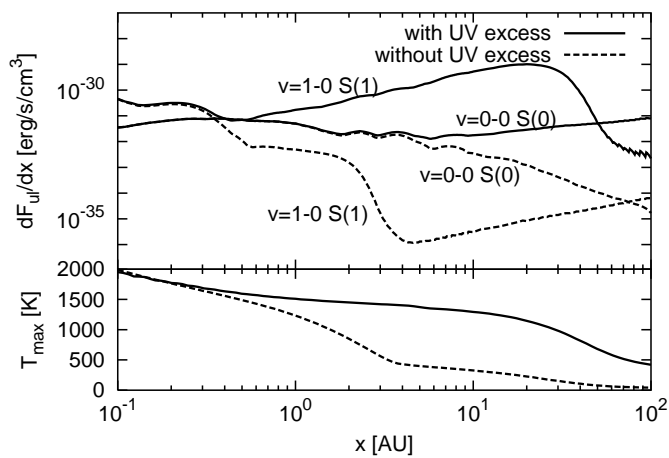}}
\caption{The radial flux distributions of the $2.12 \mu m\ v=1-0\ 
 S(1)$ line and the $28\mu m\ S(0)$ line for the models with (solid
 line) and without (dashed line) UV excess radiation (top). The
 maximum gas temperature at each radius is also plotted at the bottom. 
 If the central star has UV excess radiation, the $2.12 \mu m\ v=1-0\ 
 S(1)$ line emission comes mainly from the radius of 20 AU, while the
 $28\mu m\ S(0)$ line emission comes from all over the disk.}
\label{f4.3}
\end{figure}

Figure \ref{f4.3} represents the radial flux distributions,
$dF_{ul}/dx=(2\pi x/4\pi 
d^2)\int_{-z_{\infty}}^{z_{\infty}}\tilde{\eta}_{ul}dz$, of the $2.12
\mu m\ v=1\rightarrow 0\ S(1)$ line and the $28\mu 
m\ S(0)$ line for the models with (solid line) and without (dashed line)
UV excess radiation. The maximum gas temperature at each radius,
$T_{\rm max}(x)=\max\{T(x,z)|0<z<z_{\infty}\}$, is also plotted at the
bottom of the figure.
We can see from the figure that if the central star has a UV excess, 
the $2.12 \mu m\ v=1\rightarrow 0\ S(1)$ line emission comes mainly
from around the radius of 20 AU, which is consistent with that
derived from the observed peak-to-peak velocity separation of the
H$_2$ emission line due to the rotation velocity of the disk (Bary et
al. 2003). The line flux distribution increases with increasing radius
inside 20 AU since {the line emissivity} integrated over the vertical
direction, $\int_{-z_{\infty}}^{z_{\infty}}\tilde{\eta}_{ul}dz$, does 
not change with radius and $dF_{ul}/dx$ is almost proportional to the
radius. This is because the transition line arises from hot gas with
$T>1000$K and the maximum gas temperature of the disk satisfies this
condition. The maximum temperature does not change very much at $x<20$AU
due the weak dependence of the photoelectric heating rate on the UV
photon flux in this range (see e.g., Weingartner \& Draine 2001b). 
Meanwhile, the line flux distribution decreases suddenly beyond 20 AU
because the gas temperature falls rapidly below 1000 K due to the
stronger dependence of the photoelectric heating rate on the UV flux in
this range. 
For the model without UV excess, the line emission mainly comes from
the very inner region of the disk, $x<0.3$AU, because the UV irradiation
by the central star is not strong enough
to heat the gas at the outer disk to high enough temperatures as 
we have seen in
Sect. 2.4. The line flux distribution increases with increasing radius
at the outer disk, $x>3$AU, since the populations of the ro-vibrational
levels are controlled not by the collisional excitation but by the UV
pumping process (see Fig.~\ref{f3.2}b), and the radiation energy density
integrated over the vertical direction is almost constant to the radius
in this region.

\begin{figure}
\centering
\resizebox{\hsize}{!}{\includegraphics{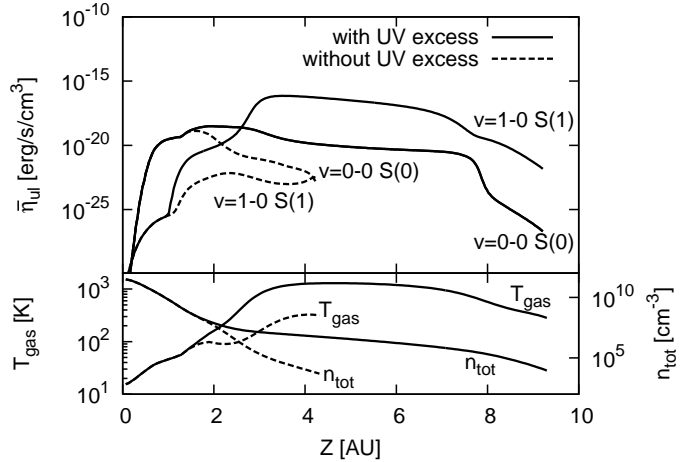}}
\caption{The vertical profiles of {the emissivity} of the
 $2.12 \mu m\ v=1-0\ S(1)$ line and the $28\mu m\ S(0)$ line at the
 radius of 10 AU for the models with (solid line) and without (dashed
 line) UV excess radiation (top). The vertical profiles of the gas
 temperature and density are also plotted at the bottom. For the model
 with UV excess radiation, the $2.12 \mu m\ v=1-0\ S(1)$ line
 emission comes from hot surface layer with $T_{\rm gas}>1000$K, while
 the $28\mu m\ S(0)$ line emission comes from cooler region near the
 midplane with $T_{\rm gas}>50$K.}
\label{f4.4}
\end{figure}
%

\begin{figure}
\centering
\resizebox{\hsize}{!}{\includegraphics{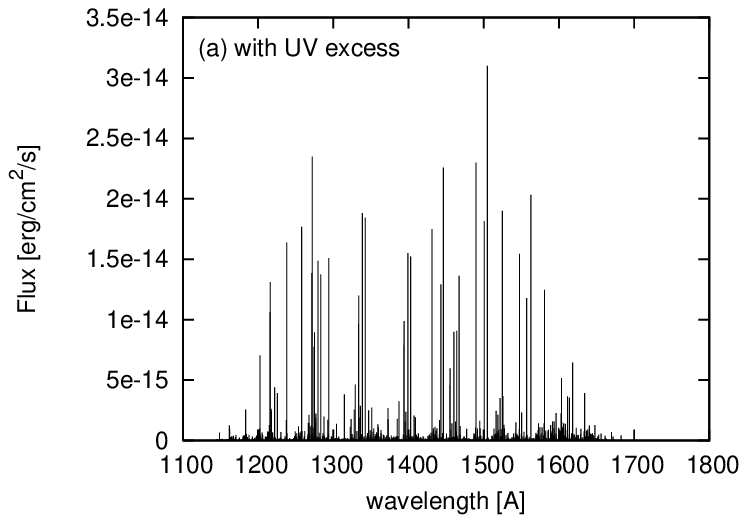}}
\resizebox{\hsize}{!}{\includegraphics{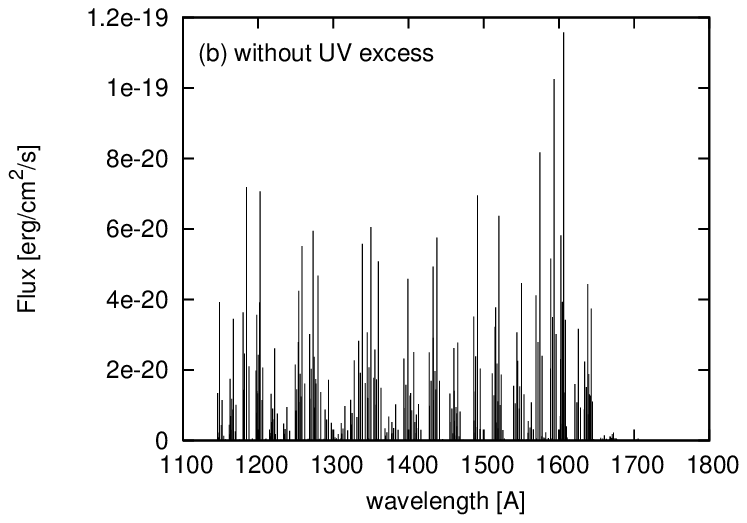}}
\caption{Same as Fig.~\ref{f4.1}, but for the ultraviolet
 ($1100{\rm \AA}<\lambda <1800{\rm \AA}$) wavelength band.}
\label{f4.5}
\end{figure}
%
\begin{figure}
\resizebox{\hsize}{!}{\includegraphics{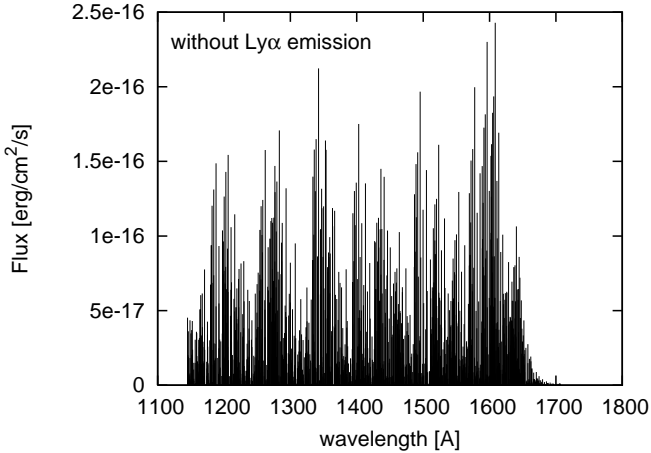}}
\caption{The ultraviolet line spectra of molecular hydrogen for the
 model without Ly$\alpha$ line emission from the central star.}
\label{f4.6}
\end{figure}
%
\begin{figure}
\resizebox{\hsize}{!}{\includegraphics{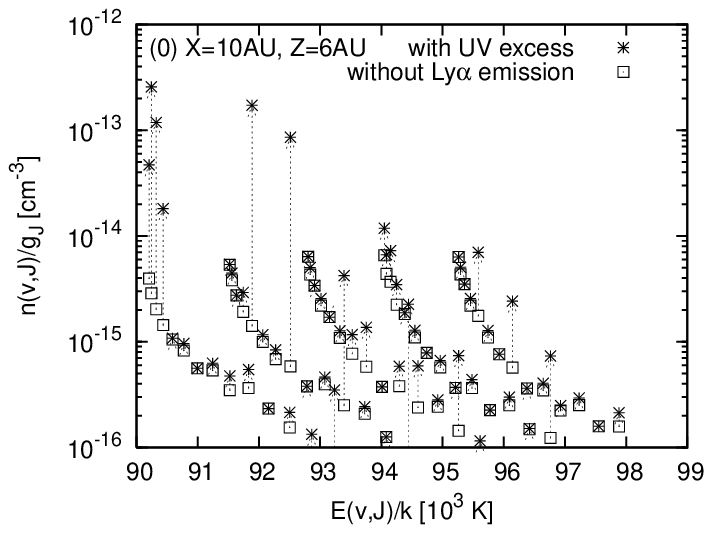}}
\resizebox{\hsize}{!}{\includegraphics{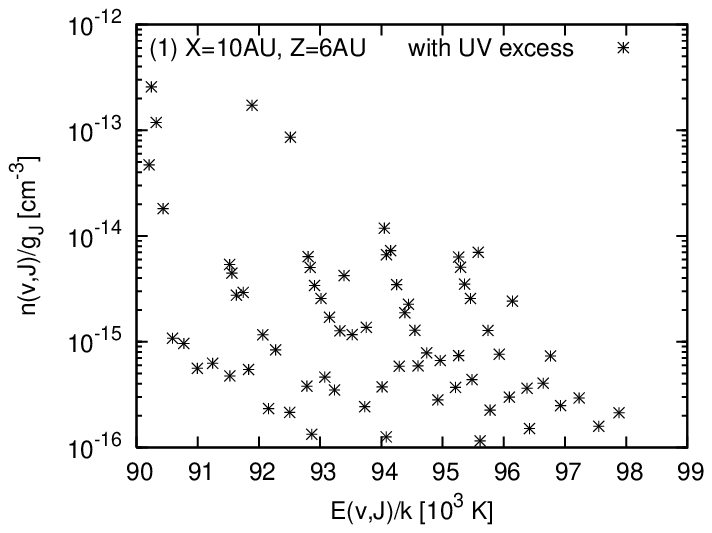}}
\resizebox{\hsize}{!}{\includegraphics{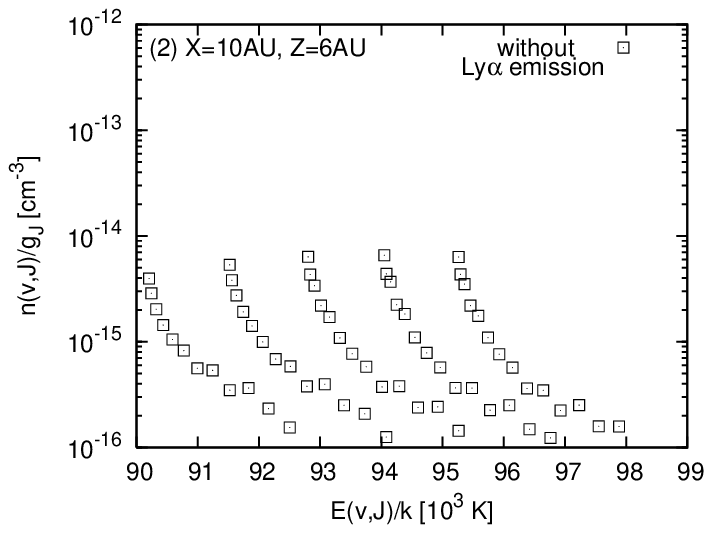}}
\caption{The level populations in the excited electronic state at
 $(X,Z)=(10{\rm AU}, 6{\rm AU})$ for the models with UV excess
 (asterisks) and without Ly$\alpha$ emission (open squares).
 Some specific levels are highly populated due to irradiation by the
 strong Ly$\alpha$ line emission from the central star.}
\label{f4.7}
\end{figure}

The $28\mu m\ S(0)$ line emission comes from a cooler region, compared
with the $2.12 \mu m\ v=1\rightarrow 0\ S(1)$ line, with the gas
temperature of $T>100$ K. If the central star has UV excess
radiation, the radial flux distribution of the line is almost constant
all over the disk, that is, {the line emissivity} integrated over
the vertical direction decreases with increasing radius. This is because
the gas is heated up to $T>100$ K even in the dense region near the
midplane in the inner disk, while it is heated up only in the less dense
region of the disk surface in the outer disk (see Fig.~\ref{f2.2}a). 
Meanwhile, for the model without UV excess, the gas temperature does
not reach 100 K at $x>20$ AU (see Fig.~\ref{f2.2}b), which makes the
line flux very weak at the outer disk. The flux
distributions for the models with and without UV excess are
identical at the inner disk, $x<3$AU, because the gas temperature
reaches 100 K even in the region where it is controlled by the dust
temperature through gas-grain collisions, and the dust temperatures are
identical in both models (see Fig.~\ref{f2.2}).

\begin{table*}
 \caption[]{The observed and calculated ultraviolet line flux of
 molecular hydrogen [erg/s/cm$^2$]}\label{T4.2} 
 $$ 
 \begin{array}{cc|cccc|cc|cccc} 
  \hline 
   \lambda & {\rm Line} & {\rm Model\ A} & {\rm Model\ B} & {\rm Model\ C} & {\rm Obs.}^{\mathrm{a}} & \lambda & {\rm Line} & {\rm Model\ A} & {\rm Model\ B} & {\rm Model\ C} & {\rm Obs.}^{\mathrm{a}} \\
  ({\rm \AA})  &  & (\times 10^{-15}) & (\times 10^{-21}) & (\times 10^{-16}) & (\times 10^{-15}) & ({\rm \AA})  &  & (\times 10^{-15}) & (\times 10^{-21}) & (\times 10^{-16}) & (\times 10^{-15}) \\
   \hline 
  1161.7 & 0-1\ R(0) &  0.16 &  4.42 &  0.08 &  &  1148.7 & 1-1\ R(3) &  0.65 &  0.04 &  0.18 & 4.6 \\
  1166.3 & 0-1\ P(2) &  0.25 &  8.74 &  0.14 &  &  1161.9 & 1-1\ P(5) &  1.25 &  0.05 &  0.30 & 10.9 \\
  1217.3 & 0-2\ R(0) &  2.61 & 13.22 &  0.37 &  &  1202.5 & 1-2\ R(3) &  7.04 &  0.08 &  0.79 & 11.3 \\
  1222.0 & 0-2\ P(2) &  4.40 & 26.14 &  0.71 &  &  1216.2 & 1-2\ P(5) & 13.11 &  0.09 &  1.15 &  \\
  1274.6 & 0-3\ R(0) &  7.75 & 23.68 &  0.70 & 27.4 &  1257.9 & 1-3\ R(3) & 17.68 &  0.06 &  0.89 & 18.1 \\
  1279.6 & 0-3\ P(2) & 14.86 & 46.82 &  1.37 & 39.2 &  1272.0 & 1-3\ P(5) & 23.49 &  0.07 &  1.12 & 20.5 \\
  1333.6 & 0-4\ R(0) &  9.64 & 28.21 &  0.84 & 42.8 &  1314.8 & 1-4\ R(3) &  3.81 &  0.01 &  0.17 & 12.2 \\
  1338.7 & 0-4\ P(2) & 18.83 & 55.75 &  1.65 & 73.1 &  1329.3 & 1-4\ P(5) &  4.61 &  0.01 &  0.21 & 7.5 \\
  1393.9 & 0-5\ R(0) &  7.94 & 23.21 &  0.69 & 35.3 &  1372.7 & 1-5\ R(3) &  2.67 &  0.01 &  0.12 & 3.2 \\
  1399.1 & 0-5\ P(2) & 15.51 & 45.88 &  1.36 & 73.8 &  1387.5 & 1-5\ P(5) &  3.23 &  0.01 &  0.15 & 7.1 \\
  1455.0 & 0-6\ R(0) &  4.59 & 13.28 &  0.40 & 20.8 &  1431.2 & 1-6\ R(3) & 17.48 &  0.06 &  0.83 & 29.0 \\
  1460.4 & 0-6\ P(2) &  9.00 & 26.23 &  0.78 & 41.6 &  1446.3 & 1-6\ P(5) & 22.57 &  0.07 &  1.04 & 44.2 \\
  1516.4 & 0-7\ R(0) &  1.78 &  5.08 &  0.15 & 21.2 &  1489.8 & 1-7\ R(3) & 23.00 &  0.08 &  1.13 & 48.2 \\
  1521.8 & 0-7\ P(2) &  3.50 & 10.03 &  0.30 & 16.2 &  1505.0 & 1-7\ P(5) & 30.99 &  0.09 &  1.44 & 57.5 \\
  1577.3 & 0-8\ R(0) &  0.42 &  1.18 &  0.04 &  &  1547.6 & 1-8\ R(3) & 15.45 &  0.05 &  0.74 & 35.3 \\
  1582.6 & 0-8\ P(2) &  0.83 &  2.33 &  0.07 &  &  1562.7 & 1-8\ P(5) & 20.33 &  0.06 &  0.94 & 37.2 \\
  1636.4 & 0-9\ R(0) &  0.05 &  0.13 &  0.00 &  &  1603.5 & 1-9\ R(3) &  5.16 &  0.02 &  0.24 & 11.2 \\
  1641.7 & 0-9\ P(2) &  0.09 &  0.26 &  0.01 &  &  1618.2 & 1-9\ P(5) &  6.44 &  0.02 &  0.29 & 11.6 \\
  1692.5 & 0-10\ R(0) &  0.00 &  0.00 &  0.00 &  &  1656.2 & 1-10\ R(3) &  0.57 &  0.00 &  0.03 &  \\
  1697.6 & 0-10\ P(2) &  0.00 &  0.01 &  0.00 &  &  1670.0 & 1-10\ P(5) &  0.69 &  0.00 &  0.03 &  \\ \hline
  1162.2 & 0-1\ R(1) &  0.17 &  1.74 &  0.09 &  &  1162.0 & 1-1\ R(6) &  0.91 &  0.00 &  0.14 &  \\
  1169.8 & 0-1\ P(3) &  0.28 &  2.55 &  0.15 &  &  1183.4 & 1-1\ P(8) &  2.55 &  0.00 &  0.24 &  \\
  1217.7 & 0-2\ R(1) &  2.43 &  5.21 &  0.52 & 9.1 &  1215.8 & 1-2\ R(6) & 10.64 &  0.00 &  0.49 &  \\
  1225.6 & 0-2\ P(3) &  3.93 &  7.64 &  0.82 &  &  1238.0 & 1-2\ P(8) & 16.37 &  0.01 &  0.56 & 11.5 \\
  1275.0 & 0-3\ R(1) &  8.94 &  9.33 &  1.12 & 24.6 &  1271.2 & 1-3\ R(6) & 13.84 &  0.00 &  0.43 & 14.1 \\
  1283.2 & 0-3\ P(3) & 13.74 & 13.70 &  1.71 & 28.0 &  1294.0 & 1-3\ P(8) & 15.09 &  0.00 &  0.46 & 13.0 \\
  1333.9 & 0-4\ R(1) & 11.98 & 11.12 &  1.40 & 7.9 &  1327.8 & 1-4\ R(6) &  2.55 &  0.00 &  0.08 & 6.1 \\
  1342.4 & 0-4\ P(3) & 18.43 & 16.32 &  2.12 & 64.9 &  1351.2 & 1-4\ P(8) &  2.72 &  0.00 &  0.08 & 2.8 \\
  1394.1 & 0-5\ R(1) &  9.89 &  9.15 &  1.15 & 52.4 &  1385.2 & 1-5\ R(6) &  1.79 &  0.00 &  0.05 &  \\
  1402.8 & 0-5\ P(3) & 15.21 & 13.44 &  1.75 & 73.1 &  1409.2 & 1-5\ P(8) &  1.91 &  0.00 &  0.06 & 2.2 \\
  1455.2 & 0-6\ R(1) &  5.97 &  5.24 &  0.68 & 30.9 &  1443.1 & 1-6\ R(6) & 12.92 &  0.00 &  0.39 & 11.3 \\
  1464.0 & 0-6\ P(3) &  9.08 &  7.68 &  1.03 & 42.1 &  1467.4 & 1-6\ P(8) & 13.62 &  0.00 &  0.42 & 17.6 \\
  1516.5 & 0-7\ R(1) &  2.45 &  2.00 &  0.27 & 21.2 &  1500.7 & 1-7\ R(6) & 18.14 &  0.00 &  0.55 & 19.7 \\
  1525.4 & 0-7\ P(3) &  3.66 &  2.94 &  0.41 & 17.9 &  1524.9 & 1-7\ P(8) & 19.01 &  0.01 &  0.59 & 23.5 \\
  1577.1 & 0-8\ R(1) &  0.60 &  0.47 &  0.07 &  &  1557.2 & 1-8\ R(6) & 11.78 &  0.00 &  0.36 & 17.0 \\
  1586.0 & 0-8\ P(3) &  0.88 &  0.68 &  0.10 &  &  1581.0 & 1-8\ P(8) & 12.46 &  0.00 &  0.38 & 17.5 \\
  1636.1 & 0-9\ R(1) &  0.07 &  0.05 &  0.01 &  &  1611.3 & 1-9\ R(6) &  3.65 &  0.00 &  0.11 &  \\
  1644.8 & 0-9\ P(3) &  0.10 &  0.08 &  0.01 &  &  1634.2 & 1-9\ P(8) &  3.92 &  0.00 &  0.12 & 5.5 \\
  1692.0 & 0-10\ R(1) &  0.00 &  0.00 &  0.00 &  &  1661.5 & 1-10\ R(6) &  0.39 &  0.00 &  0.01 &  \\
  1700.3 & 0-10\ P(3) &  0.00 &  0.00 &  0.00 &  &  1682.8 & 1-10\ P(8) &  0.42 &  0.00 &  0.01 &  \\
   \hline 
 \end{array}
 $$ 
 \begin{list}{}{}
  \item[$^{\mathrm{a}}$] Observations by Herczeg et al. (2002).
 \end{list}
\end{table*}

Furthermore, the vertical profiles of {the emissivity},
$\tilde{\eta}_{ul}$ (Eq. [\ref{eq.4-5}]), of the $2.12 \mu m\
v=1\rightarrow 0\ S(1)$ line and the $28\mu m\ S(0)$ line at 10 AU for
the models with (solid line) and without (dashed line) UV excess
radiation are plotted in Fig.~\ref{f4.4}. We can see from the figure
that in the surface layer of the disk {the emissivity} of
both lines for both models basically increases with decreasing vertical
height due to the increasing gas density. For the model with UV
excess radiation, {the emissivity} of the $2.12 \mu m\
v=1\rightarrow 0\ S(1)$ line decreases suddenly where the gas
temperature falls below $\sim 1000$ K. 
The corresponding disk height is $z\sim 3.5$AU$\sim 17H$, where
$H=\cs/\Omega$ is the disk scale height. The optical depth from the disk
surface to $z\sim 3.5$AU is small enough, $\tau_{ul}\sim 1.5\times
10^{-3}$, for the line.  
Meanwhile, in the model without
UV excess, it is independent of the gas temperature because the
level populations are controlled by the UV pumping process (see
Sect. 3.2), and it decreases near the midplane where the UV irradiation
cannot penetrate. 
{The emissivity} of the $28\mu m\ S(0)$
line has a peak where the gas temperature drops below 100K in both models.
The corresponding disk height is $z\sim 1.9$AU$\sim 9.5H$, where
the line is also optically thin ($\tau_{ul}\sim 6.6\times 10^{-3}$) for
the model with UV excess. 
{The emissivity} in the far cooler region close to the midplane is
still relatively large due to the high gas density, but it decreases
distinctly at $T<50$ K. Therefore, although the gas in the disk is
concentrated near the midplane, these lines come only from the upper
region of the disk. Also, the figure indicates that the line flux traces
the gas density and temperature where the gas temperature falls below
the critical value.

\subsection{The ultraviolet spectra}

The resulting line spectra in the ultraviolet wavelength band are
plotted in Figs.~\ref{f4.5}a and \ref{f4.5}b for the models with and
without UV excess radiation, respectively. If the central star has
a UV excess, some of those lines, whose upper levels are in excited
electronic states and populated via the UV pumping process, 
become stronger by more than six orders
of magnitude than those in the model without UV excess.

\begin{figure}
\centering
\resizebox{\hsize}{!}{\includegraphics{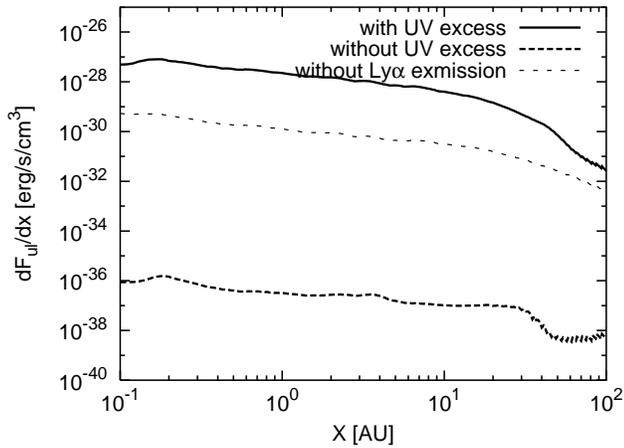}}
\caption{The radial flux distributions of the 1272.0A 
 $v=1\rightarrow 3\ P(5)$ line for the models with (solid  line) and
 without (dashed line) UV excess radiation, and without
 Ly$\alpha$ emission (dotted line). The radiation flux decreases with
 increasing radius for all models because the UV irradiation flux,
 which pumps hydrogen molecules in the ground electronic state, decreases.} 
\label{f4.8}
\end{figure}

Some of the strongest line fluxes we calculate in the model with UV
excess (Model A) are listed in Table~\ref{T4.2} and compared with
observations by the HST and the FUSE (Herczeg et al. 2002). The
calculated strong line fluxes result from the excitation due to the
strong Ly$\alpha$ line emission from the central star and are consistent
with the observations.
The line fluxes in the model without UV excess (Model B) are also
presented in the table for comparison. In addition,
in order to see the effect of pumping by Ly$\alpha$ emission, we
calculate the level populations and UV line fluxes by neglecting 
Ly$\alpha$ emission (Model C). The gas density and temperature profiles
for the model with UV excess are used here. The resulting line
emission is presented in Fig.~\ref{f4.6} and Table~\ref{T4.2}, and is
much weaker than the observed values.

The level populations in the excited electronic
state at the radius of 10 AU and the vertical height of 6 AU
for Model A (asterisks) and C (open squares) are plotted in
Fig.~\ref{f4.7}. {Figures labeled (1) and (2) show the level
populations for Model A and C, respectively, and they are plotted
together in figure (0) for comparison. The dotted lines in figure (0)
connect the level populations of Model A and C in the same energy
levels. The figures} clearly show 
that specific level populations are extremely high for Model A 
(with UV excess radiation) due to the irradiation of the Ly$\alpha$
emission from the central star. The transitions from these levels result
in the strong line emission in Fig.~\ref{f4.5}a.
Meanwhile, the level populations for Model C (without Ly$\alpha$
emission) are low and distributed smoothly since they are excited only
by continuum radiation. This is why the line fluxes are weak and many
lines have similar strength in Fig.~\ref{f4.6}. 

In the case of Model B (without UV excess) the line spectra are much
weaker (Fig.~\ref{f4.5}b) due to the weak UV irradiation from the
central star and the low level populations of pre-excited hydrogen
molecules in the ground electronic state (Fig.~\ref{f3.2}b, see also
below). Also, the line spectra are relatively scattered, although they
result from the pumping by continuum irradiation.
This is because molecular hydrogen in highly excited electronic states
originates mainly from highly populated low energy levels ($v=0$) in the
ground electronic state, whose populations decrease suddenly with
increasing level energies due to the low temperature for the model
without UV excess (Fig.~\ref{f3.2}b). Thus, the level populations in
the excited electronic state also have steeper distribution, which
leads to the sparse line spectra. 

\begin{figure}
\centering
\resizebox{\hsize}{!}{\includegraphics{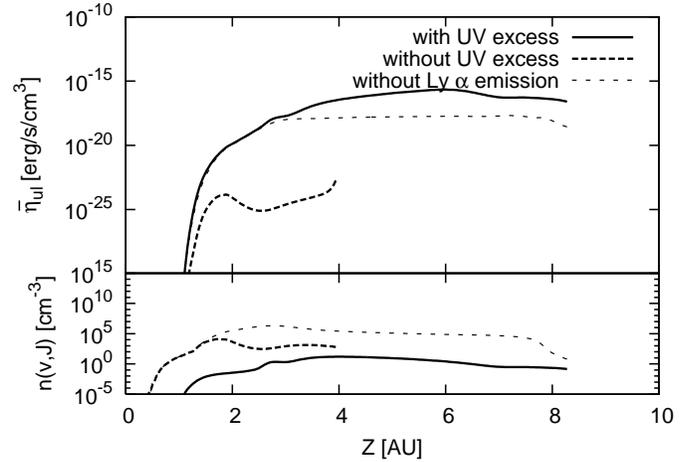}}
\caption{The vertical profiles of {the emissivity of the 1272.0A 
 $v=1\rightarrow 3\ P(5)$ line} at the
 radius of 10 AU for the models with (solid line) and without (dashed
 line) UV excess radiation, and without Ly$\alpha$ emission
 (dotted line)(top). The energy density profiles are similar to
 the number density profiles of H$_2$ in the pre-excited level in
 the ground electronic state (bottom).}
\label{f4.9}
\end{figure}

Figure \ref{f4.8} shows the radial flux distribution of the 1272.0\AA\ 
$v=1\rightarrow 3\ P(5)$ line, which is one of the strongest lines in
Model A, for the models with (solid line) and
without (dashed line) UV excess radiation, and without
Ly$\alpha$ emission (dotted line). The flux decreases with increasing
radius for all models because the line intensity is
proportional to the population of the upper transition level in the
excited electronic state, namely, the UV irradiation flux from the
central star which is
proportional to the inverse squares of the radius at the disk surface.

The vertical profiles of {the line emissivity}
at 10AU for these models are also plotted in Fig.~\ref{f4.9} (top). 
At the disk surface, the profiles of $\tilde{\eta}_{ul}$ are similar to
those of the number densities of pre-excited hydrogen molecules in the
ground electronic state, $n(v,J)$, plotted in the bottom of
Fig.~\ref{f4.9}. This is because {the line emissivity} is
proportional to the number density of H$_2$ in the 
upper transition level of the line, $n(v^*=1,J^*=4)$ (the symbol $^*$
means that it is in the excited electronic state), which is populated
via the UV pumping process and proportional to both $n(v,J)$ and the
UV irradiation flux, $F_{\nu(vJ\rightarrow v^*J^*),
R}+F_{\nu(vJ\rightarrow v^*J^*), z}$ (Eqs. [\ref{eq.2-6}] and
[\ref{eq.2-7}]). 
In the models without UV excess and without Ly$\alpha$ emission,
the level $(v^*=1,J^*=4)$ is populated mainly from the highly populated
low energy level, $(v=0,J=3)$. Meanwhile, in the model with UV excess, the
transition from the level $(v=2,J=5)$ dominates in populating the level
$(v^*=1,J^*=4)$ because the excitation wavelength, 
$\lambda=hc/E(v=2,J=5\rightarrow v^*=1,J^*=4)=1216.2$\AA, is close to
the central wavelength of
the Ly$\alpha$ line emission. So, the number density of $n(v=0,J=3)$
is plotted for the models without UV excess and without Ly$\alpha$
emission, while that of $n(v=2,J=5)$ is plotted for the model with UV
excess in the bottom of Fig.~\ref{f4.9}.
{The line emissivity} decreases with decreasing vertical height for all
models especially near the midplane because the irradiating UV photons
are self-absorbed (used for the pumping) and can not penetrate deep in
the disk. Thus, the molecular hydrogen emission from near the midplane
is not observable in the ultraviolet region.

\section{Summary}

We have, for the first time, investigated molecular hydrogen emission from 
protoplanetary disks, taking into account the global physical structure of 
the disk and
the detailed analysis of level populations of molecular hydrogen.

First, we obtained in a self-consistent manner the density and 
temperature profiles of gas and dust, taking into account the
irradiation by a central T Tauri star. As a result, we found that
if the central star has UV excess emission over the photospheric blackbody
radiation, the gas temperature at the disk surface can reach about
1,000K even at the radius of 10 AU due to grain photoelectric heating
induced by the UV photons from the central star. 

Next, making use of the physical structure of the disk, we calculated the
level populations of molecular hydrogen in its ground electronic state
on the assumption of statistical equilibrium. The resulting level
populations tend to be in LTE, although the populations in the
upper levels become very large due to the high gas temperature if the
central star has UV excess radiation. 

Furthermore, using these level populations, we computed molecular hydrogen 
emission in the near- and mid-infrared and ultraviolet wavelength bands by 
solving the radiative
transfer equation. Consequently, we found strong line emission spectra if
the central star has UV excess radiation. The infrared line
spectra are strong because the populations of the upper levels become
large due to the high gas temperature, while the ultraviolet line
emission is intense because the strong UV irradiation, including that due to 
Ly$\alpha$, by the central
star pumps the hydrogen molecule from the ground to the excited
electronic states. We compared the results of our calculation with
the observations to find that the calculated near-infrared intensity of
the $v=1\rightarrow 0\ S(1)$ line is in good agreement with that observed
towards TW Hya, as are the ultraviolet lines. 
Our predictions 
for the pure rotational line intensities are much below those observed by 
Thi et al. (2001b) but this may be related to the large ISO beam used.

\begin{acknowledgements}
We would like to thank an anonymous referee for his/her comments which
 improved our paper. We are also grateful to Dr. R. Wagenblast for
 giving us the numerical code for molecular hydrogen level populations.
Astrophysics at UMIST and the University of Manchester is supported by a
 grant from PPARC. Also, H.N. is supported by ``The 21st Century COE
 Program of Origin and Evolution of Planetary Systems" in MEXT and
 the Research Fellowships of the JSPS 16036205 and 15540237.
\end{acknowledgements}

\appendix

\section{Thermal processes}

The following three processes which are dominant under conditions in
protoplanetary disks are considered as heating and cooling
processes in this paper. 

\subsection{Grain photoelectric heating}
The photoelectric emission from dust grains induced by far ultraviolet
(FUV) photons from the central star dominates the heating at the
surface layer of protoplanetary disks (see Sect. 2.4), especially if the
star has UV excess radiation (Sect. 2.2 and Appendix C). 
In this paper we adopt the photoelectric heating rate calculated by
Weingartner \& Draine (2001b), 
\begin{equation}
\Gamma_{\rm pe}=1.0\times 10^{-26}\epsilon G_{\rm FUV}n_{\rm tot} {\rm ergs\ cm}^{-3}{\rm s}^{-1},
\end{equation}
\begin{equation}
\epsilon=\dfrac{C_0+C_1T^{C_4}}{1+C_2(G_{\rm FUV}\sqrt{T}/n_e)^{C_5}[1+C_3(G_{\rm FUV}\sqrt{T}/n_e)^{C_6}]},
\end{equation}
where $G_{\rm FUV}$ represents the FUV fields measured in unit of the
equivalent average interstellar radiation flux of $1.6\times 10^{-3}$
ergs cm$^{-2}$ s$^{-1}$ (Habing 1968), and $T$, $n_{\rm tot}$, and $n_e$
represent the gas temperature, the number densities of hydrogen nuclei
and electrons, respectively. The parameter set of $C_0$ - $C_6$ is taken
from Weingartner \& Draine (2001b), where we choose the dust size
distribution model for dense clouds with the ratio of visual extinction
to reddening of $R_V\equiv 
A(V)/E(B-V)=5.5$, the total C abundance per H nucleus in the log-normal
population of $b_{\rm C}=3.0\times 10^{-5}$, and the grain volumes 
assumed to be the same as those of the diffuse interstellar medium (Case
B of WD01a; see also Appendix D).

\subsection{Gas-grain collisions}
The energy exchange between gas and dust particles through 
collisions is one of the most important thermal processes in
protoplanetary disks in which the density is high enough. We use the
following cooling rate (heating rate if $T_d>T$) in this paper,
\begin{equation}
\Lambda_{\rm gr}=3.5\times 10^{-34}n_{\rm tot}^2T^{0.5}(T-T_d) {\rm ergs\ cm}^{-3}{\rm s}^{-1}
\end{equation}
(Burke \& Hollenbach 1983; Tielens \& Hollenbach 1985). The dust
temperature $T_d$ is obtained on the assumption of the local radiative
equilibrium in Sect. 2.1.

\subsection{Radiative cooling}
Radiative transitions among the fine-structure levels of OI (63$\mu$m)
and CII (158$\mu$m), and the rotational line transitions of CO contribute to
cooling gas in the surface layer of the disks where the gas density is low
enough. The cooling rate due to the transition
from upper level $i$ to lower level $j$ of species $s$ is calculated as
\begin{displaymath}
\Lambda_s(\nu_{ij})=h\nu_{ij}\beta_{\rm esc}(\tau_{ij})
\end{displaymath}
\begin{equation}
\hspace*{1cm}\times[n_i\{A_{ij}+B_{ij}P(\nu_{ij})\}-n_jB_{ji}P(\nu_{ij})],
\end{equation}
where $n_i$, $A_{ij}$, $B_{ij}$, and $h\nu_{ij}$ are the population
density of level $i$, the Einstein probabilities for spontaneous and
stimulated emission, and the energy difference between levels $i$ and
$j$, respectively (e.g., de Jong et al. 1980).
As the photon escape probability, $\beta_{\rm esc}(\tau_{ij})$, we use
the approximate function given by de Jong et al. (1980).
Assuming that the nearest boundary from a point $(r,z)$ is the disk
surface $(r,z_{\infty})$, we approximately calculate the optical depth
averaged over the line, $\tau_{ij}$, as
\begin{eqnarray}
\tau_{ij}(z) &=& \dfrac{A_{ij}c^3}{8\pi\nu_{ij}^3}\int_z^{z_{\infty}}n_i(z')\biggl[\dfrac{n_j(z')g_i}{n_i(z')g_j}-1\biggr]\dfrac{dz'}{\delta v_d(z')} \\
 &\approx & \dfrac{A_{ij}c^3}{8\pi\nu_{ij}^3}\dfrac{n_i(z)}{n_{\rm H}(z)}\biggl[\dfrac{n_j(z)g_i}{n_i(z)g_j}-1\biggr]\dfrac{N_{\rm H}(z)}{\delta v_d(z)},
\end{eqnarray}
where $g_i$ is the statistical weight factors of the level $i$,
$\delta v_d(z)=\{2kT(z)/\mu m\}^{0.5}$ is the Doppler line width, and
$N_{\rm H}(z)=\int_z^{z_{\infty}}n_{\rm H}(z')dz'$ is the column 
density of hydrogen nuclei from $z$ to $z_{\infty}$.
The background radiation $P(\nu_{ij})$ is calculated by assuming that it
is contributed by the radiation from the central star and the infrared
dust emission, 
\begin{displaymath}
P(\nu_{ij}) = (1/8)(R/R_*)^2B(\nu_{ij}, T_*)\exp[-\tau_{d,R}(\nu_{ij})]
\end{displaymath}
\begin{equation}
\hspace*{1.5cm} +(1/2)B(\nu_{ij}, T_d)\{1-\exp[-\tau_{d,z}(\nu_{ij})]\},
\end{equation}
where $T_*$ and $R_*$ are the temperature and the radius of the central
star, respectively, and $\tau_{d,R}(\nu_{ij})=\int_{R_*}^R
\kappa_{\nu_{ij}}\rho dR$ and $\tau_{d,z}(\nu_{ij})=\int_z^{z_{\infty}}
\kappa_{\nu_{ij}}\rho dz$ are the optical depth in the radial and the
vertical directions, respectively. The dust temperature $T_d$ is obtained
on the assumption of local radiative equilibrium, as discussed in Sect. 2.1. 

The level populations of CII and OI are obtained by solving the equations
of statistical equilibrium,
\begin{equation}
n_i\sum_{j\ne i}R_{ij}=\sum_{j\ne i}n_jR_{ji},
\end{equation}
where
\begin{equation}
\begin{array}{ll}
R_{ij}=A_{ij}+B_{ij}P(\nu_{ij})+C_{ij}, & i>j, \\
R_{ij}=B_{ij}P(\nu_{ij})+C_{ij}, & i<j.
\end{array}
\end{equation}
The symbol $C_{ij}$ represents the collisional transition rate. The
collisional de-excitation rate coefficients are taken from Tielens \&
Hollenbach (1985), and the collisional excitation rates
are calculated as $C_{ij}=C_{ji}(g_j/g_i)\exp(h\nu_{ij}/kT) (i<j)$.
The level population of CO is assumed to be in LTE, which will
overestimate the CO cooling rate when $n_{\rm H}<10^7$ cm$^{-3}$
(cf. Kamp \& van Zaldelhoff 2001). The abundances of OI, CII, and CO are
calculated as is described in Appendix B.2.

\section{Chemistry}

\subsection{Hydrogen photoionization}

It is known that extreme UV photons with energy of $h\nu>13.6$ eV are
mainly consumed to photoionize atomic hydrogen. Thus, we simply consider
that most hydrogen atoms are ionized if
\begin{equation}
\int_{R_*}^r\alpha_{\rm H} n_{\rm tot}^2dr < \int_{\nu_{\rm EUV}}^{\infty}\dfrac{F_{\nu, R}}{h\nu}d\nu,
\end{equation}
where $\alpha_{\rm H}=2.58\times 10^{-13}$ cm$^3$ s$^{-1}$ is the
recombination coefficient for atomic hydrogen at $10^4$ K, $F_{\nu,R}$
is the radiation flux from the central star in Eq. (\ref{eq.2-6}), and
$h\nu_{\rm EUV}\equiv 13.6$ eV (e.g., Osterbrock 1989;
Hollenbach et al. 1994). The temperature of this ionized region is
simply assumed to be $T=10^4$K, since we are interested in the structure
of non-ionized region and the temperature of this ionized surface layer
does not affect the global structure of the disks very much.

\subsection{Carbon and oxygen chemistry}
In order to calculate the radiative cooling by C$^+$, O and CO, we treat
a very simple chemistry for carbon and oxygen, and calculate the
abundances of these species. 
The total fractional abundances of carbon
and oxygen with respect to hydrogen nuclei are assumed to be $x({\rm C}_{\rm
tot})=x({\rm C}^+)+x({\rm C})+x({\rm CO})=7.86\times 10^{-5}$ and $x({\rm
O}_{\rm tot})=x({\rm O})+x({\rm CO})=1.8\times 10^{-4}$, respectively.
The number density of carbon monoxide is calculated
using the following two chemical equilibrium:
\begin{equation}
k_0n({\rm C}^+)n({\rm H}_2)=k_1n({\rm CH}_x)n({\rm O})+G_{\rm FUV}\Gamma_{{\rm CH}_x}n({\rm CH}_2)
\end{equation}
and
\begin{equation}
k_1n({\rm CH}_x)n({\rm O})=G_{\rm FUV}\Gamma_{\rm CO}n({\rm CO}),
\end{equation}
as
\begin{eqnarray}
n({\rm CO}) &=& \dfrac{k_1n({\rm CH}_x)n({\rm O})}{G_{\rm FUV}\Gamma_{\rm CO}} \\
 &=& \dfrac{k_1n({\rm O})}{G_{\rm FUV}\Gamma_{\rm CO}}\dfrac{k_0n({\rm C}^+)n({\rm H}_2)}{k_1n({\rm O})+G_{\rm FUV}\Gamma_{{\rm CH}_x}n({\rm CH}_2)},
\end{eqnarray}
following Nelson \& Langer (1997) (cf. Langer 1976), where $k_0=5\times
10^{-16}$ cm$^3$ s$^{-1}$ is the rate coefficient of the radiative
association reaction C$^+ +$H$_2\rightarrow$CH$_2^+ + h\nu$ and
$k_1=5\times 10^{-10}$cm$^3$ s$^{-1}$ is the rate coefficient of the
reaction between the hydrocarbon radicals and atomic oxygen to form CO.
The symbol $G_{\rm FUV}$ represents the UV radiation described in Sect. 2.2,
and $\Gamma_{{\rm CH}_x}=5\times 10^{-10}$ s$^{-1}$ and $\Gamma_{\rm
CO}=10^{-10}$ s$^{-1}$ are the photodissociation rates of the hydrocarbon
radicals and the carbon monoxide, respectively.
More detailed chemistry should be solved in future (e.g., Aikawa et
al. 2002; Markwick et al. 2002; Kamp \& Dullemond 2004).

\section{Central stellar radiation}

Here we model the radiation from the T Tauri star by means of
black body emission at an effective stellar temperature plus optically
thin hydrogenic thermal bremsstrahlung emission at a higher temperature
(e.g., Lago et al. 1984; Costa et al. 2000) and Ly $\alpha$ line
emission (e.g., Hergzeg et al. 2002).
The specific luminosity at {a wavelength $\lambda$} of the
photospheric black body radiation is calculated as
\begin{equation}
L_{\lambda, {\rm bb}}=4\pi^2 R_*^2 B_{\lambda}(T_*), \label{eq.C-1}
\end{equation}
while the luminosity of the coronal hydrogenic thermal bremsstrahlung
emission is computed as
\begin{equation}
L_{\lambda, {\rm br}}\approx \eta_{\lambda, {\rm br}}V, \label{eq.C-2}
\end{equation}
where $V$ is the volume of the bremsstrahlung emission region and
$\eta_{\lambda, {\rm br}}$ is the emissivity,
\begin{displaymath}
\eta_{\lambda, {\rm br}}=2.0\times 10^{-36}n_e^2T_{\rm br}^{-1/2}\lambda^{-2}
\end{displaymath}
\begin{equation}
\hspace*{1.5cm}\times\exp(-hc/\lambda kT_{\rm br})\ \ {\rm erg\ s}^{-1}{\rm cm}^{-3}{\rm \AA}^{-1}, \label{eq.C-3}
\end{equation}
(e.g., Rybicki \& Lightman 1979). Here $n_e$, $T_{\rm br}$, and $c$ are
the electron number density, the temperature of the bremsstrahlung
emission, and the light speed, respectively. In addition, we take into
account the luminosity of the Ly 
$\alpha$ line emission at the stellar surface, 
\begin{equation}
L_{\lambda, {\rm Ly}\alpha}=L_{\lambda_0, {\rm Ly}\alpha}\exp\{-(\lambda-\lambda_0)^2/\sigma^2\}, \label{eq.C-4}
\end{equation}
where we simply assume that the line has a Gaussian profile with the
central wavelength of $\lambda_0$, the peak luminosity of $L_{\lambda_0, {\rm
Ly}\alpha}$,
and the line width of $\sigma$. 

\begin{figure}
\resizebox{\hsize}{!}{\includegraphics{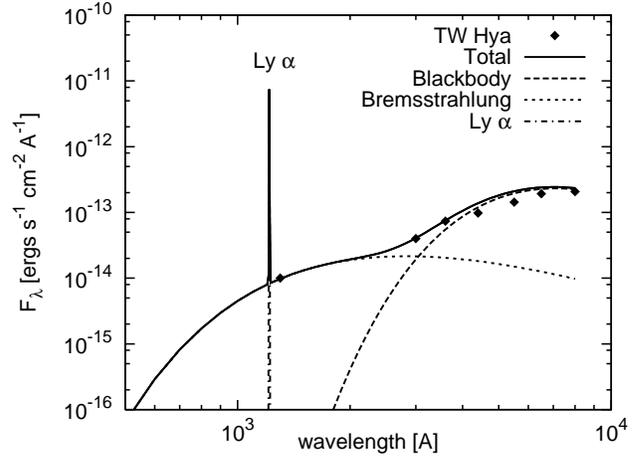}}
\caption{The radiation flux of the central T Tauri star (solid line)
 modeled by black body radiation (dashed line), thermal bremsstrahlung
 emission (dotted line), and Ly $\alpha$ line emission (dot-dashed
 line). The filled diamonds show the observational data towards TW Hya.}
\label{fC.1}
\end{figure}

Now, we choose the physical parameters in
Eqs. (\ref{eq.C-2})-(\ref{eq.C-4}) by comparing the calculated
{radiation flux} to observations towards TW Hya.
The effective stellar temperature and radius of TW Hya of $T_*=4000$K
and $R_*=1R_{\odot}$, and the distance to TW Hya of $d=56$ pc are
adopted in the calculation. 
Figure~\ref{fC.1} shows the calculated {radiation
flux density} and the observational data.
The dashed, dotted, dot-dashed, and solid lines represent the black
body, the bremsstrahlung, the Ly $\alpha$ line, and the total radiation,
respectively. The filled diamonds are the observations. 
The observational data at the I, R, V, B, and U bands are taken from
Herbst et al. (1994). The median value of 11.02 mag is adopted for the V
magnitude. The data of 2.0 $\times 10^{-14}$ and 1.0 $\times 10^{-14}$
erg cm$^{-2}$ s$^{-1}$ \AA$^{-1}$ at the wavelengths of 1300\AA\ and 3000\AA,
respectively, are measured from the observation by Costa et al. (2000). 
The best fit physical parameters of $T_{\rm br}=2.5\times 10^4$K and
$n_e^2V= 3.68\times 10^{56} $cm$^{-3}$ are obtained for the thermal
bremsstrahlung emission.
We note that 
according to the analysis of the IUE data towards pre-main-sequence
stars by Johns-Krull, Valenti, \& Linsky (2000), many classical T Tauri
stars have color temperatures of $\sim 10^4$ K, derived from the mean
continuum flux at the wavelengths of 1958\AA\ and 1760\AA, independent of
the effective stellar temperatures.
For the Ly $\alpha$ line emission, we adopt the central wavelength of
$\lambda_0=1215.67$\AA, the luminosity ratio of the line peak to the
continuum radiation of $L_{\lambda_0, {\rm Ly}\alpha}/(L_{\lambda_0, {\rm
bb}}+L_{\lambda_0, {\rm br}})=10^3$, and the line width of $\sigma=2.01$\AA,
following Herczeg et al. (2002; see also Ardila et al. 2002).

The radiation flux from the central star is calculated as $F_{\lambda,
{\rm star}}=(L_{\lambda, {\rm bb}}+L_{\lambda, {\rm br}}+L_{\lambda,
{\rm Ly}\alpha})/(4\pi R_*^2)$ and $F_{\lambda, {\rm star}}=L_{\lambda,
{\rm bb}}/(4\pi R_*^2)$ for the models with and without UV excess
radiation, respectively. {We note that $F_{\nu, {\rm star}}$ in
Sect. 2.2 is related to $F_{\lambda, {\rm star}}$ as $\lambda
F_{\lambda, {\rm star}}=\nu F_{\nu, {\rm star}}$, where $\nu$ is the
frequency.} 
In this paper we adopt the physical parameters of $T_*$, $T_{\rm br}$,
$\lambda_0$, $L_{\lambda_0, {\rm Ly}\alpha}/(L_{\lambda_0, {\rm
bb}}+L_{\lambda_0, {\rm br}})$, and $\sigma$ as mentioned above, and set
the stellar 
radius to be 2$R_{\odot}$ for calculating the radiation from the central
star, that is, we take the luminosity of the central star ($L_*=4\pi
R_*^2F_*$) to be four times larger than that of TW Hya.

\section{Grain opacity}

In order to calculate the absorption ($\kappa_{\nu}$) and scattering
($\sigma_{\nu}$) coefficients of dust grains, we use the following dust
model in this paper. First, as the dust components we adopt silicate and
carboneous grains, both of which are considered as components of 
interstellar dust, and water ice, which is expected to be
formed in the cold and dense region of protoplanetary disks.
The optical properties of the carboneous grains are assumed to have a
continuous distribution of graphite-like properties for larger sizes
and PAH-like properties in the small size limit, following Li \& Draine
(2001). The mass fractional abundances of the above species are taken to
be consistent with the solar elemental abundances: $\zeta_{\rm
sil}=0.0043$, $\zeta_{\rm carbon}=0.0030$, and $\zeta_{\rm ice}=0.0094$ 
(Anders \& Grevesse 1989). Their bulk densities are set to be $\rho_{\rm
sil}=3.5$, $\rho_{\rm graphite}=2.24$, and $\rho_{\rm ice}=0.92$ g
cm$^{-3}$ (see Li \& Draine 2001 for the PAH's bulk density). Their
sublimation temperatures are simply assumed to be $T_{\rm sil}=1500$K,
$T_{\rm carbon}=2300$K, and $T_{\rm ice}=150$K (e.g., Adams \& Shu 1986).

\begin{figure}
\resizebox{\hsize}{!}{\includegraphics{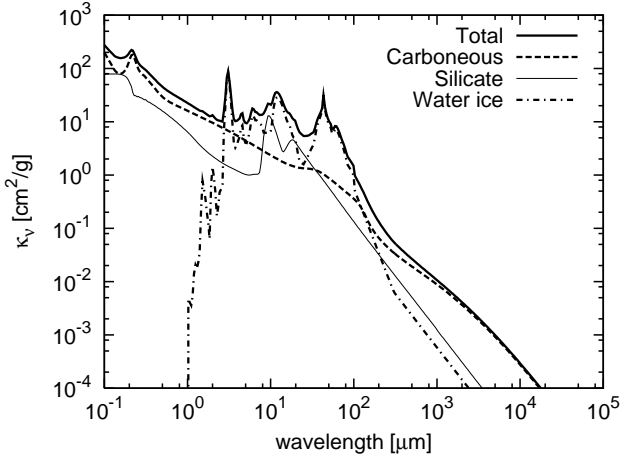}}
\caption{The monochromatic absorption coefficient (thick solid line)
 of dust grains consisting of carboneous (dashed line), silicate (thin
 solid line), and water ice (dot-dashed line).}
\label{fD.1}
\end{figure}

We assume that the silicate and carboneous grain particles have the size 
distribution obtained by WD01a, which can
reproduce the observational extinction curve of dense clouds with the
ratio of visual extinction to reddening $R_V\equiv A(V)/E(B-V)=5.5$
(see also Cardelli, Clayton \& Mathis 1989). We use the model with the
parameter of $b_{\rm C}=3.0\times 10^{-5}$, which represents the total C
abundance per H nucleus in the log-normal population, and the grain
volumes are assumed to be the same as those of the diffuse interstellar
medium (Case B of WD01a). The water ice is simply assumed to have the MRN
size distribution of $dn/da\propto a^{-3.5}$, where $a$ is the radius of
the dust particles (Mathis, Rumpl, \& Nordsieck 1977).

The total mass absorption coefficient $\kappa_{\nu}$ and the total mass
scattering coefficient $\sigma_{\nu}$ are calculated as 
\begin{equation}
\{\kappa_{\nu}, \sigma_{\nu}\}=\sum_s\dfrac{\int (dn/da)_sa^3\{\kappa_{\nu, s}(a), \sigma_{\nu, s}(a)\}da}{\int(dn/da)_s a^3da},
\end{equation}
where $s$ represents the dust component - silicate, carboneous, or
water ice. The mass absorption and scattering coefficients of dust
particles with radius $a$ is given by
\begin{equation}
\{\kappa_{\nu, s}(a), \sigma_{\nu, s}(a)\}=\dfrac{3}{4a\rho_s}\{Q_{\rm abs}(a,\nu), Q_{\rm sca}(a,\nu)\}\zeta_s
\end{equation}
(e.g.,  Miyake \& Nakagawa 1993). We adopt the absorption factors $Q_{\rm
abs}(a,\nu)$ described in Li \& Draine (2001) for PAH molecules.
The other absorption and scattering efficiency
factors, $Q_{\rm abs}(a,\nu)$ and $Q_{\rm sca}(a,\nu)$, are computed
by means of the Mie theory (Bohren \& Huffman 1983), simply assuming spherical
grains. In order to calculate the efficiency factors, we use the
dielectric function of the `smoothed UV astronomical silicate' and 
graphite by Draine \& Lee (1984), Laor \& Draine (1993), and WD01a
(http://www.astro.prinston.edu/\textasciitilde draine). Also we use the
refractive indices by Miyake \& Nakagawa (1993, see also references
therein) for the water ice. 

The calculated monochromatic absorption coefficient is shown by the
thick solid line in Fig.~\ref{fD.1}. Each component of carboneous,
silicate, and water ice is also plotted in the dashed, thin solid, and
dot-dashed lines, respectively.

\end{document}